\documentclass[11pt,a4paper]{article}
\usepackage{jheppub}
\DeclareMathOperator{\tr}{tr}

\newcommand{\Z}{\mathbb{Z}}

\newcommand{\1}{\mbox{1}\hspace{-0.25em}\mbox{l}}

\title{\boldmath Higher spin AdS$_3$ holography with extended supersymmetry}

\author[a]{Thomas Creutzig,}
\author[b]{Yasuaki Hikida}
\author[c]{and Peter B. R\o nne}

\affiliation[a]{Department of Mathematical and Statistical Sciences, 632 CAB, University of Alberta, Edmonton, Alberta T6G 2G1, Canada}
\affiliation[b]{Department of Physics, Rikkyo University, 3-34-1 Nishi-Ikebukuro, Toshima, Tokyo 171-8501, Japan}
\affiliation[c]{University of Luxembourg, Mathematics Research Unit, FSTC,
Campus Kirchberg, 6, rue Coudenhove-Kalergi, L-1359 Luxembourg-Kirchberg, Luxembourg}

\emailAdd{creutzig@ualberta.ca}
\emailAdd{hikida@rikkyo.ac.jp}
\emailAdd{peter.roenne@uni.lu}

\abstract{We propose a holographic duality between a higher spin AdS$_3$
gravity with so$(p)$ extended supersymmetry and a large $N$ limit of a 2-dimensional Grassmannian-like model with
a specific critical level $k=N$ and a non-diagonal modular invariant.
As evidence, we show the match of one-loop partition functions.
Moreover, we construct symmetry generators of the coset model for low spins which are dual
to gauge fields in the supergravity. Further, we discuss a possible relation to superstring theory
by noticing an ${\cal N}=3$ supersymmetry of critical level model at finite $k,N$.
In particular, we examine BPS states and marginal deformations. Inspired by the supergravity side, we also propose and test another large $N$ CFT dual obtained as a $\Z_2$ automorphism truncation of a similar coset model, but at a non-critical level. }


\arxivnumber{1406.1521}

\begin{document}

\maketitle
\flushbottom

\section{Introduction}

Higher spin gauge theory is believed to be related to the massless limit of
superstring theory, and it should be useful for understanding typical properties of
stringy states.
A non-trivial higher spin gauge theory is given by Vasiliev theory \cite{Vasiliev:2003ev} which has been
used to construct simplified versions of the AdS/CFT correspondence.
For example, 4d Vasiliev theory was proposed to be dual to the 3d O$(N)$ vector model \cite{Sezgin:2002rt,Klebanov:2002ja}, and it was conjectured in \cite{Gaberdiel:2010pz,Gaberdiel:2012uj} that
3d  higher spin gravity theory in \cite{Prokushkin:1998bq} is related to a large $N$ limit of the 2d $W_N$ minimal model.
Recently, an extended version of 4d Vasiliev theory was proposed \cite{Chang:2012kt} to be dual to the Aharony-Bergman-Jafferis (ABJ) theory
\cite{Aharony:2008ug,Aharony:2008gk}, which is known to be dual to superstring theory on AdS$_4 \times \mathbb{C}$P$^3$.
Therefore we have now a relation (dubbed ABJ triality in \cite{Chang:2012kt}) between 4d Vasiliev theory, ABJ theory and superstring theory.
The construction of the lower dimensional version should be useful to understand the
triality better since, in general, a more detailed analysis is possible in lower dimensional models.
Several works have already appeared in this direction
\cite{Gaberdiel:2013vva,Creutzig:2013tja,Ahn:2013oya,Candu:2013fta,Gaberdiel:2014yla,Beccaria:2014jra}, however the relation
between 3d Vasiliev theory and superstring theory is not as clear as for the ABJ triality.
One of the aims of this paper is to take a further step to uncover this relation.

In our previous work \cite{Creutzig:2013tja}, we have proposed that the ${\cal N}=2$ higher spin supergravity
with $M \times M$ matrix valued fields (or equivalently with U$(M)$ Chan-Paton factor) in \cite{Prokushkin:1998bq} is
dual to an 't Hooft limit of
\begin{align}
 \frac{\text{su}(N+M)_k \oplus \text{so}(2NM)_1}{\text{su}(N)_{k+M} \oplus \text{u}(1)_\kappa}
\label{coset}
\end{align}
with $\kappa = NM(N+M)(N+M+k)$. The limit is taken with large $N,k$, but finite $M$
and finite 't Hooft parameter
\begin{align}
 \lambda = \frac{N}{N+M+k} \, .
\label{thooft}
\end{align}
The 't Hooft parameter is identified with the mass parameter of the matter fields in the supergravity.
The duality in the case where $M=2$ coincides with the one proposed in \cite{Gaberdiel:2013vva}.

We would like to relate the duality to superstring theory.
For this purpose we may restrict ourselves to a singlet sector of the U$(M)$ Chan-Paton factor with large $M$ as in the ABJ triality \cite{Chang:2012kt}.
Thus it is natural to consider the ${\cal N}=2$
Grassmannian Kazama-Suzuki coset  \cite{Kazama:1988uz,Kazama:1988qp}
\begin{align}
 \frac{\text{su}(N+M)_k \oplus \text{so}(2NM)_1}{\text{su}(N)_{k+M} \oplus
\text{su}(M)_{k+N} \oplus \text{u}(1)_\kappa}
\label{KScoset}
\end{align}
for large $M$. However, we do not know what the string theory dual to this Kazama-Suzuki model is.
On the other hand, the coset \eqref{coset} with $M=2$ is based on a Wolf space and
is known to have (non-linear) ${\cal N}=4$ superconformal symmetry \cite{Spindel:1988sr,VanProeyen:1989me,Sevrin:1989ce}.%
\footnote{It was already suggested in \cite{Henneaux:2012ny} that coset models based on
Wolf spaces could be used to construct higher spin holography with large supersymmetry.}
It was argued in \cite{Gaberdiel:2013vva} that the target space of the involved superstring theory can be
identified as AdS$_3 \times $S$^3 \times$S$^3 \times$S$^1$
due to the large ${\cal N}=4$ supersymmetry, see also \cite{Gaberdiel:2014yla,Beccaria:2014jra}.%
\footnote{An interesting connection to the alternating spin chain of \cite{OhlssonSax:2011ms} was mentioned in \cite{Gaberdiel_talk}.}
Earlier works on the large ${\cal N}=4$ holography may be found in \cite{Elitzur:1998mm,deBoer:1999rh,Gukov:2004ym}.
However, in their approach, we do not know how to introduce the large $M$ structure of the U$(M)$
Chan-Paton factor.

In order to improve this situation, we would like to take advantage of the enhanced supersymmetry in the bulk. In general, the Vasiliev theory \cite{Prokushkin:1998bq} is a one parameter family of
${\cal N}=2$ supergravities with
matrix valued fields, which includes higher spin gauge fields and massive matter
parameterized by $\lambda$. However at the specific value $\lambda = 1/2$,
where some matter fields become massless, the field content can consistently be truncated to the half to obtain an ${\cal N}=p $ enhanced supersymmetry.
The symmetry algebra has also been studied in \cite{Henneaux:2012ny}.
Let us set $p = 2n $ with $n = 0,1,2,\ldots$.
We will see that the supersymmetry can be extended to ${\cal N}=p +1$ and the superalgebra
$\text{osp}(p + 1|2)$ is generated by (see eq. (10.16) in \cite{Prokushkin:1998bq})
\begin{align}
 T_{\alpha \beta} =\{ y_\alpha , y_\beta \} \, , \quad
 Q_{\alpha}^I = y_\alpha \otimes \phi^I \, , \quad M^{IJ} = [ \phi^I , \phi^J ] \, .
 \label{ospi}
\end{align}
Here $y_\alpha$ $(\alpha =1,2)$ and $\phi^I$ $(I=1,2,\ldots ,p+1)$ satisfy
\begin{align}
 [y_\alpha , y_\beta] = 2 i \epsilon_{\alpha \beta} \, , \quad
\{ \phi^I , \phi^J \} = 2 \delta^{IJ} \, .
\label{yphi}
\end{align}
The Clifford elements $\phi^I$ generate the Clifford algebra $C_{p+1}$, which can be
realized by $2^{n} \times 2^{n}$ matrices.

In this paper we propose that the dual model is given by
\begin{align}
 \frac{\text{su}(N+M)_k}{\text{su}(N)_{k} \oplus \text{u}(1)_{kNM(N+M)}}
\label{bcoset}
\end{align}
with $k=N$, but large $N$ and $M = 2^n$.%
\footnote{We consider the case where $M$ is of the form $M= 2^n$ with $n=0,1,2,\cdots$ since we are interested in the case with extended supersymmetry. However,  we can easily extend the duality to the case with generic integer $M$.}
We consider a non-diagonal modular invariant such that the su$(N)_N$ factor in the denominator
could be expressed by free fermions in the adjoint representations of su$(N)$.
For ${\cal N}=1$ (or $n=0$) the duality is the same as the one proposed in \cite{Beccaria:2013wqa}
if a level-rank duality as in \cite{Bowcock:1988vs,Altschuler:1988mg} is applied.
In the dual form, it is known that the bosonic model actually has ${\cal N}=1$ supersymmetry
assuming the non-diagonal modular invariant \cite{Goddard:1986ee,Douglas:1987cv,Hornfeck:1990zw,Ahn:1990nr,Schoutens:1990xg}.
See also \cite{Gopakumar:2012gd,Ahn:2012bp,Ahn:2013ota,Isachenkov:2014zua} for studies of this in the context of higher spin holography. We show that the one-loop partition function of the coset model \eqref{bcoset} in the large $N$ limit with $k=N$  reproduces that of the dual gravity theory. Moreover, we construct symmetry generators of the coset model explicitly
at low spins which should be dual to low spin gauge fields in the dual gravity theory.

It can be shown that the supersymmetric coset \eqref{coset}%
\footnote{The coset model \eqref{coset} actually does not preserve any supersymmetry.
Even so we call this model as the supersymmetric coset in order to distinguish this coset and the bosonic
coset \eqref{bcoset}. Anyway, by  gauging the su$(M)$ factor in \eqref{coset}, the model becomes the
${\cal N}=2$ supersymmetric Kazama-Suzuki model \eqref{KScoset}.}
 with $k=N\pm M$ and $M=2^{n-1}$
is directly related to the bosonic coset \eqref{bcoset} by decoupling some free fermions.
The supersymmetric form of the coset is more
suitable in the discussion on the possible relation to superstring theory.
In fact, we find that the Grassmannian Kazama-Suzuki model in \eqref{KScoset} with $k=N+M$
has enhanced ${\cal N}=3$ superconformal symmetry even for finite $N$, and this is also one
of our main results. The large superconformal symmetry restricts the possible target space of the dual superstring theory to a large extent, and currently only a few candidates are known
\cite{Yamaguchi:1999gb,Argurio:2000tg,Argurio:2000xm}.
In this way, we find a possibility to construct a three dimensional version of the ABJ triality by circumventing the
problems that previously existed.%
\footnote{We can see from the argument in the ABJ triality \cite{Chang:2012kt} that a higher spin theory  with both U($M$) Chan-Paton factor and U($M$) invariant condition has string-like spectrum. However, our coset may not have enough supersymmetry such as to make the string-like theory to really be the superstring theory contrary to the ABJ triality.
Therefore we cannot deny the possibility that there is no superstring theory dual to our coset.}
We examine the duality between the Kazama-Suzuki model \eqref{KScoset} with $k=N+M$ and
a superstring theory by comparing BPS states and marginal deformations.

This paper is organized as follows:
In section \ref{gravity1}, we introduce the higher spin gravity with ${\cal N}=p+1$ enhanced supersymmetry.
We summarize its spectrum and obtain the one-loop partition function.
In section \ref{bcm} we study the duality between the higher spin gravity in section \ref{gravity1} and
the bosonic coset \eqref{bcoset} with $k=N$ and large $N$. We reproduce the gravity partition function from the limit of the model. We also construct low spin generators, which should be dual to low spin gauge fields in the bulk.
In section \ref{scm} we examine the supersymmetric models \eqref{coset} and \eqref{KScoset}.
The partition function of the model \eqref{coset} with $k=N-M$ and large $N$ is shown to
reproduce the gravity partition function after some fermions are decoupled.
Furthermore, we  show the Kazama-Suzuki model \eqref{KScoset}
with $k=N+M$ has ${\cal N}=3$ superconformal symmetry at the critical level,
and utilizing the fact we study relations between this model and a superstring theory.

In section \ref{gravity2} the higher spin algebra of the bulk side is constructed as a truncation via an $\Z_2$ automorphism of the ${\cal N}=2$ higher spin supergravity
with $M \times M$ matrix valued fields in \cite{Prokushkin:1998bq}. For $M=2^{p/2}$ the truncated higher spin algebra is shown to have an osp$(p+1|2)$ subalgebra indicating a so$(p+1)$ extended supersymmetry algebra on the boundary at the linear level. In section \ref{sec:orbi} we consider the CFT side again, but this time we show that the Grassmannian coset CFT dual to the untruncated higher spin supergravity at $\lambda=1/2$ contains an $\Z_2$ automorphism working like the $\Z_2$ automorphism on the bulk. This CFT has a non-critical level in contrast to the previous models, but the level is chosen such that the automorphism is given by level-rank duality composed with charge conjugation. We then perform the truncation according to this $\Z_2$ automorphism on the CFT side to get a third candidate for the dual CFT (at least in the large $N$ limit). We make some checks of OPEs and matter states of the duality between this naive orbifold CFT and the bulk theory.

Finally, we conclude this work and comment on open problems in section \ref{conclusion}.
Two technical appendices then follow; The detailed analysis on the CFT partition
function is given  in appendix \ref{PF}, and several techniques to examine the symmetry algebra of
coset models are collected in appendix \ref{sec:sgcm}.

\section{Gravity partition function}
\label{gravity1}

We consider the higher spin gauge theory on AdS$_3$ with ${\cal N}= 2n+1$ supersymmetry as
in \cite{Prokushkin:1998bq,Henneaux:2012ny},
which can be obtained by a $\Z_2$-truncation of the ${\cal N}=2$ higher spin supergravity with
$M \times M$ $(M=2^n)$ matrix valued fields, as mentioned in the introduction.
In this section we introduce the spectrum of the higher spin gravity and obtain
its one-loop partition function.
See also section \ref{gravity2} for more details including an analysis of the supersymmetry.

The theory has a gauge sector and a matter sector. The gauge sector includes
spin $s=1,2,3, \ldots$ bosonic higher spin gauge fields and spin $s=3/2,5/2,7/2, \ldots$ fermionic
higher spin gauge fields. These fields take values in a u$(M)$ Lie algebra with the exception
that the trace part of u$(M)$ decouples for the spin 1 gauge field.
The matter sector includes two complex $M \times M$ matrix valued scalar fields with
mass squared $m^2 = - 3/4$ and two massless $M \times M$ matrix valued Dirac fermions.
Choosing the condition at the boundary of AdS properly, the conformal dimensions of the
dual operators are $(h , \bar h)= (1/4,1/4)$ and $(3/4,3/4)$ for each complex scalar,
and $(h , \bar h ) = (1/4,3/4)$ and $(3/4 ,1/4)$ for the Dirac fermions.

We parametrize the modulus of the torus at the AdS boundary by $q$.
Then the one-loop partition function of a bosonic spin $s \, (=2,3,4,\ldots)$ gauge field on AdS$_3$
is given by \cite{Gaberdiel:2010ar}
\begin{align}
 Z^{(s)}_B (q) = \prod_{n=s}^\infty \frac{1}{|1 - q^n|^2} \, .
\end{align}
It was shown in \cite{Creutzig:2011fe} that the same expression also holds for a spin 1 gauge field.
For a fermionic spin $s-1/2 \,  (=3/2 ,5/2 , 7/2 , \ldots)$  gauge field, the one-loop partition function is \cite{Creutzig:2011fe}
\begin{align}
 Z^{(s)}_F  (q) = \prod_{n=s}^{\infty}  |1 + q^{n-1/2} |^2 \, .
\end{align}
Therefore, the contribution to the one-loop partition function from the gauge sector is
\begin{align}
 Z_\text{gauge} (q) =
 \left( \prod_{s=2}^\infty  Z^{(s)}_B (q) Z^{(s)}_F (q) \right)^{M^2} \left( Z^{(1)}_B (q) \right )^{M^2 - 1} \, ,
\end{align}
which follows from the spectrum mentioned above.

For  a complex scalar with dual conformal dimension $(h,h)$, the one-loop partition function is \cite{Giombi:2008vd,David:2009xg}
\begin{align}
 Z^{h}_\text{scalar} (q) = \prod_{m,n = 0}^\infty \frac{1}{(1 - q^{h +m} \bar q^{h +n})^{2}} \, ,
\end{align}
and for a Dirac fermion with dual conformal dimension $(h , h - 1/2)$, $(h -1/2 ,h)$ it is
\cite{Creutzig:2011fe}
\begin{align}
 Z^h_\text{spinor} (q) = \prod_{n,m=0}^\infty
  \left( 1 + q^{h +m} \bar q^{h + n - 1/2} \right) \left( 1 + q^{h +m - 1/2} \bar q^{h + n}\right) \, .
\end{align}
The total contribution from the matter sector is
\begin{align}
 Z_\text{matter} (q) =
  \left( Z^{1/4}_\text{scalar} (q) ( Z^{3/4}_\text{spinor} (q) )^2 Z^{3/4}_\text{scalar} (q) \right)^{M^2} \, ,
  \label{matterpf}
\end{align}
which leads to the total partition function
\begin{align}
 Z_\text{bulk} (q) = Z_\text{gauge} (q) Z_\text{matter} (q) \, .
 \label{gpf}
\end{align}

In order to compare this expression to the CFT partition function,
it is convenient to rewrite it in terms of supercharacters of Young diagrams
\cite{Gaberdiel:2011zw,Candu:2012jq,Creutzig:2013tja,Candu:2013fta}.
Here we introduce  a supercharacter
\begin{align}
 \text{sch}_\alpha ({\cal U}_h) = \sum_{T \in \text{STab}_\alpha} \prod_{j \in T} q^{h+j/2} \, ,
\quad
( {\cal U}_h )_{jj} = (-1)^j q^{h+j/2} \, ,
\label{schU}
\end{align}
where $\text{STab}_\alpha$ is the super Young tableau of shape $\alpha$ (see, e.g., \cite{Candu:2012jq} for a
detailed explanation).  Then we can define
\begin{align}
 {\cal B}^{h,M}_\alpha (q) = \sum_{\alpha_1 , \ldots , \alpha_M}
 c^{\alpha}_{\alpha_1 \ldots \alpha_{M}} \prod_{A = 1}^M \text{sch}_{\alpha_A} ({\cal U}_h)
 \label{sBhM}
\end{align}
with
\begin{align}
 c^{\alpha}_{\alpha_1 \ldots \alpha_{M}}
 = \sum_{\beta_1, \ldots , \beta_{M-2}} c^\alpha_{\alpha_1  \beta_1}
\left( \prod_{A=1}^{M-3} c^{\beta_A}_{\alpha_{A+1} \beta_{A+1}}\right)
 c^{\beta_{M-2}}_{\alpha_{M-1} \alpha_M}
\label{cM}
\end{align}
by using the Clebsch-Gordan coefficients $c^\alpha_{\beta \gamma}$ of $\text{gl}(\infty)_+$ which were introduced in \cite{Candu:2012jq}.
In this language, the gravity partition function \eqref{gpf} takes the form
\begin{align}\label{eq:partrewritten}
 Z_\text{bulk} (q) = Z_\text{gauge} (q) \sum_{\Lambda^l , \Lambda^r} |{\cal B}^{1/4,M}_{\Lambda^l}(q) {\cal B}^{1/4,M}_{\Lambda^r} (q)|^2 \, .
\end{align}
We now want to calculate this from the CFT side of the duality.

\section{The bosonic coset models}
\label{bcm}

In this section we study the coset \eqref{bcoset} with $k=N $ and large $N$ but finite $M \, (= 2^n)$, and
examine its duality to the higher spin gravity with ${\cal N}=2n+1$ extended supersymmetry.
In the next subsection we define the coset model in more detail.
Furthermore, we compute the partition function  of the coset model in the large $N$ limit,
and see the match with the gravity partition function \eqref{gpf}.
In sections \ref{bgenerator} we investigate the symmetry of the coset model.
It will turn out to be difficult to construct symmetry generators directly in the form of the coset \eqref{bcoset}
with $k=N$.
So we utilize a (conjectured) map from the coset with $k=N$ to the coset with $k=N+M$ and
some fermions decoupled as in \eqref{eq:compboscoset}, where the latter model is easier to
analyze its symmetry generators.

\subsection{Spectrum and partition function}
\label{bpf}

The coset \eqref{bcoset} is a  bosonic model, so one may wonder how it is possible that the coset
is dual to a supersymmetric gravity theory.
In fact, we need to utilize a mechanism of supersymmetry enhancement for the purpose.
The bosonic coset \eqref{bcoset} may be obtained by adding a su$(M)$ factor to
the Grassmannian model
\begin{align}
 \frac{\text{su}(N+M)_k}{\text{su}(N)_{k} \oplus \text{su}(M)_{k}\oplus \text{u}(1)_{kNM(N+M)}}
\label{gcoset}
\end{align}
as discussed in \cite{Creutzig:2013tja}.
This bosonic  coset is known to be dual to the following coset \cite{Bowcock:1988vs,Altschuler:1988mg}
\begin{align}
 \frac{\text{su}(k)_N \oplus \text{su}(k)_M}{\text{su}(k)_{N+M}} \, .
\label{dcoset}
\end{align}
A special thing happens at $N=k$ since $\text{su}(N)_N$ at this critical level has a description in terms of free fermions in the adjoint representation of $\text{su}(N)$.
Moreover, it was argued in \cite{Goddard:1986ee,Douglas:1987cv} that the model has
${\cal N}=1$ supersymmetry.
For $M=1$, the coset with $N=k$ has the ${\cal N}=1$ supersymmetric $W_k$ as symmetry algebra,
of which the bosonic $W_k$ algebra is a sub-algebra \cite{Hornfeck:1990zw,Ahn:1990nr,Schoutens:1990xg}.
This fact was used to construct the duality with the higher spin gravity with ${\cal N}=1$
supersymmetry (i.e. $M=1$ or equivalently $n=0$) in \cite{Beccaria:2013wqa}.%
\footnote{The higher spin supergravity used in \cite{Creutzig:2012ar} also has ${\cal N}=1$ supersymmetry,
but it is different from the ${\cal N}=1$ theory discussed here. For instance, the bosonic gauge fields
in the higher spin theory used in  \cite{Creutzig:2012ar} have only even spin $s=2,4,6,\ldots$, but the
${\cal N}=1$ supergravity used in \cite{Beccaria:2013wqa} has gauge fields of spin $s=2,3,4,\ldots$.}

In order to define the model, we also need to specify the spectrum leading to a modular
invariant partition function. In  \cite{Schoutens:1990xg} a non-diagonal modular
invariant is chosen such that adjoint fermions from su$(N)_N$ in the numerator of
\eqref{dcoset} act like generators of symmetry. The adjoint free fermions can be
expressed by so$(N^2 -1)$ current algebra with level one.
The states of the coset \eqref{dcoset}
with $M=1$ are labeled by $(\Lambda_+; \Lambda_-)$, where
$\Lambda_+$ is the highest weight for su$(N)_N$ and $\Lambda_-$ for su$(N)_{N+1}$.
At large $N$, it is convenient to express $\Lambda_\pm$ by two Young diagrams
$(\Lambda^l_\pm , \Lambda^r_\pm)$, see, e.g., \cite{Candu:2012jq}.
Then, as explained in \cite{Beccaria:2013wqa}, $\Lambda_+$ is restricted now to
$\Lambda_+ \in \Omega=\{(\Lambda^l_+ , \Lambda^r_+)|\Lambda_+^l = (\Lambda_+^r)^t\}$,
where $\alpha^t$ is the transpose of $\alpha$.
The Hilbert space is thus \cite{Beccaria:2013wqa}
\begin{align}
 {\cal H} = \bigoplus_{\Lambda_{-}}  {\cal H}_{\Lambda_{-}}
\otimes \bar {\cal H}_{\Lambda_{-}^*}
\, , \quad
 {\cal H}_{\Lambda_{-}} = \bigoplus_{\Lambda_+ \in \Omega} (\Lambda_{+};\Lambda_-) \, .
\label{sWspe}
\end{align}

We can now consider the original coset \eqref{bcoset} for general $M$. The states of that coset are labeled by
$(\Lambda_{M+N};\Lambda_N,m)$, where $\Lambda_L$ are highest weights of
$\text{su}(L)$ and $m \in \mathbb{Z}_{kNM(N+M)}$.
At large $N$, $m$ is fixed as
\begin{align}
m = N|\Lambda_{N+M}|_- - (N+M)|\Lambda_N|_-
\label{mfix}
\end{align}
where $|\alpha|$ is the number of boxes in a Young diagram $\alpha$
and $|\Lambda_L|_- = |\Lambda_L^l| - |\Lambda_L^r|$ \cite{Candu:2012jq,Creutzig:2013tja,Candu:2013fta}.
Thus the states are labeled by $(\Lambda_{M+N};\Lambda_N)$ in the 't Hooft limit.
As in \eqref{sWspe}, we consider the following spectrum
\begin{align}
 {\cal H} = \bigoplus_{\Lambda_{N+M}}  {\cal H}_{\Lambda_{N+M}}
\otimes \bar {\cal H}_{\Lambda_{N+M}^*}
\, , \quad
 {\cal H}_{\Lambda_{N+M}} = \bigoplus_{\Lambda_N \in \Omega} (\Lambda_{N+M};\Lambda_N) \, ,
\label{bspectrum}
\end{align}
where the sum in $\Lambda_N \in \Omega$ is over $\Lambda^l_N = (\Lambda^r_N)^t$.
It is easy to check that the difference of conformal weights of holomorphic and
anti-holomorphic parts is integer or half-integer $h - \bar h \in \frac12 {\mathbb Z}$.

Based on the spectrum \eqref{bspectrum}, we compute the coset partition function
in the large $N$ limit and compare it with the gravity partition function \eqref{gpf}.
In the 't Hooft limit, the character of $(\Lambda ; \Xi)$ has been computed as \cite{Creutzig:2013tja}
(see also \cite{Candu:2013fta})
\begin{align}
 b^{\lambda,M}_{\Lambda ; \Xi} (q) = X^M_0(q) q^{\frac{1 - \lambda}{2} (|\Lambda^l| + |\Lambda^r| - |\Xi^l| - |\Xi^r| )}
\sum_{\Phi , \Psi} R^{(N)}_{\Lambda \Phi} C^{(N)\Psi^*}_{\Phi \Xi^*} B^{1,M}_{\Psi^l} (q) B^{1,M}_{\Psi^r} (q) \, .
\end{align}
The vacuum character with $(\Lambda ; \Xi) = (0,0)$ is given by%
\footnote{Since we compare one-loop partition functions, we neglect the tree level contribution
$q^{- c/24}$, where $c$ is the central charge.}
\begin{align}
 b^{\lambda , M}_{0;0} (q) =
 X^M_0 (q) = \left( \prod_{s=2}^\infty z_{B}^{(s)} (q) \right)^{M^2 }
 \left(z_{B}^{(1)}(q)\right)^{M^2 - 1} \, , \quad
z_B^{(s)} (q) = \prod_{n=s}^{\infty}\frac{1}{1 - q^n} \, .
\end{align}
The restriction coefficients $R^{(N)}_{\Lambda \Phi}$ and the Clebsch-Gordan coefficients
$C^{(N)\Psi}_{\Phi \Xi}$ are introduced as
\begin{align}
 \text{ch}_{\Lambda}^{N+M} (\imath (v)) = \sum_{\Phi} R^{(N)}_{\Lambda \Phi}
 \text{ch}_{\Phi}^{N+M} (v) \, ,  \quad
 \text{ch}_{\Phi}^{N} (v)  \text{ch}_{\Xi}^{N} (v)=
 \sum_{\Psi}C^{(N)\Psi}_{\Phi \Xi} \text{ch}_{\Psi}^{N} (v)
\label{restriction}
\end{align}
with $\imath(v)$ being an embedding of $\text{SU}(N)$ into $\text{SU}(N+M)$
and $ \text{ch}_{\Lambda}^{L}$ a $\text{SU}(L)$ character of representation $\Lambda$.
The function $B^{h,M}_\alpha (q)$ is the bosonic counter part of \eqref{sBhM},
which is defined as
\begin{align}
 B^{h,M}_\alpha (q) = \sum_{\alpha_1 , \ldots , \alpha_M}
 c^{\alpha}_{\alpha_1 \ldots \alpha_{M}} \prod_{A} \text{ch}_{\alpha_A} (U_h)
 \label{BhM}
\end{align}
with \eqref{cM}.
Instead of supercharacters in \eqref{schU}, the following characters are used
\begin{align}
 \text{ch}_\alpha (U_h) = \sum_{T \in \text{Tab}_\alpha} \prod_{j \in T} q^{h+j} \, ,
\quad
( U_h )_{jj} = q^{h+j} \, ,
\label{chU}
\end{align}
where $\text{Tab}_\alpha$ as the Young tableau of shape $\alpha$.

In the current case, we need to set $\lambda = 1/2$ and sum over
$\Xi \in \Omega$ as
\begin{align}
 \chi^M_\Lambda (q) = \sum_{\Xi \in \Omega} b^{1/2,M}_{\Lambda ; \Xi} (q)
\label{chiLambdaM0}
\end{align}
as  in \eqref{bspectrum}. As shown in appendix \ref{PF}, this expression can be
simplified as
 \begin{align}
 \chi_\Lambda^M (q) = \chi^M_0 (q) {\cal B}^{1/4,M}_{\Lambda^l} (q)
{\cal B}^{1/4,M}_{\Lambda^r} (q)
\label{chiLambdaM}
\end{align}
by extending the method used in \cite{Beccaria:2013wqa}.
The vacuum character is
\begin{align}
 \chi_0^M (q)  =
  \left( \prod_{s=2}^\infty z_{B}^{(s)} (q) z_{F}^{(s)} (q) \right)^{M^2 }
 \left(z_{B}^{(1)}(q)\right)^{M^2 - 1} \, , \quad
z_F^{(s)} (q) = \prod_{n =s}^\infty (1 + q^{n-1/2}) \, ,
\label{chiM0}
\end{align}
which is consistent with the spin content of higher spin gauge theory introduced in section \ref{gravity1}.
Using that $(\Lambda^l,\Lambda^r)^*=(\Lambda^r,\Lambda^l)$, we conclude that the one-loop partition function of the bosonic coset \eqref{bcoset} with the spectrum \eqref{bspectrum}
in the 't Hooft limit is given by
\begin{align}
{\cal Z}_\text{CFT} (q) = \left| \chi^M_0 (q) \right| ^2 \sum_{\Lambda^l , \Lambda^r} \left|  {\cal B}^{1/4,M}_{\Lambda^l} (q)
{\cal B}^{1/4,M}_{\Lambda^r} (q) \right|^2 \, .
\label{eq:CFTpt}
\end{align}
This reproduces the one-loop partition function for ${\cal N}=2n+1$ higher spin
supergravity with $M = 2^{n}$ in \eqref{gpf} when one compares with the rewritten form in \eqref{eq:partrewritten}.

\subsection{Symmetry generators}
\label{bgenerator}

We now construct the symmetry generators of the coset \eqref{bcoset} explicitly at low spins.
Let us consider a generic coset $\hat g/\hat h$. Then the symmetry generators are made from
currents of $\hat g$, which should not have any singular OPEs with currents of $\hat h$
(see, e.g., \cite{Bais:1987zk}).  In addition to this, we demand that the generators are primary with respect to
the energy momentum tensor. In the coset \eqref{dcoset} with $N=k$ and with the spectrum \eqref{sWspe}, the states may be generated from $(0 , \Lambda_-)$ by the action of adjoint free fermions.
Therefore, in order to construct generators of the symmetry algebra, we need to use the
adjoint free fermions along with currents of $\hat g$
\cite{Hornfeck:1990zw,Ahn:1990nr}, see also \cite{Ahn:2012bp,Ahn:2013ota}
and appendix \ref{App:Ahn}.
There the generators of super $W$-algebras at low spins are constructed
explicitly in the coset language.

The application of this method to our case does not seem to be straightforward.
This is because the critical level su$(N)_N$ model appears in the denominator of the coset, and thus the
adjoint free fermions cannot be used for the purpose to construct symmetry generators.
{}There is, however, a trick. Let us instead consider the case with $k=N+M$ in the coset \eqref{bcoset} and then
construct the symmetry algebra. We then notice that a critical level $\text{su}(N+M)_{N+M}$ model appears
in the numerator, and it can be described by free fermions in the adjoint representation of
$\text{su}(N+M)$.
It turns out that the symmetry generators then contain fields with spin $1/2$,
and we decouple these free fermions as in \cite{Goddard:1988wv}.
We will argue that the model after this decoupling actually is
directly related to the bosonic coset \eqref{bcoset} with $k=N$, which is exactly what we wanted to describe, i.e.
\begin{align}\label{eq:compboscoset}
   \frac{\text{su}(N+M)_N}{\text{su}(N)_{N} \oplus \text{u}(1)_{N^2M(N+M)}}\approx \frac{\text{su}(N+M)_{N+M}}{\text{su}(N)_{N+M} \oplus \text{u}(1)_{NM(N+M)^2}}\Big|_{\text{Free fermions decoupled}} \, .
\end{align}
Below we show this by using the description of su$(N+M)_{N+M}$ current algebra with su$(N+M)$ adjoint fermions, and we explain what the relation means in some details.
With the description by the right hand side, we construct the low spin generators of the left hand side explicitly.

\subsubsection{Decoupling free fermions}

We thus consider the coset \eqref{bcoset} with $k=N+M$ and the following Hilbert space as
\begin{align}
 {\cal H} = \bigoplus_{\Lambda_{N}}  {\cal H}_{\Lambda_{N}}
\otimes \bar {\cal H}_{\Lambda_{N}^*}
\, , \quad
 {\cal H}_{\Lambda_{N}} = \bigoplus_{\Lambda_{N+M} \in \Omega} (\Lambda_{N+M};\Lambda_N) \, .
\label{bspe}
\end{align}
We decompose su$(N+M)$ as%
\footnote{Here we use $L$ and $\bar L$ as the fundamental and the anti-fundamental representations of su$(L)$, respectively.}
\begin{align}
 \text{su}(N+M) = \text{su}(N) \oplus \text{su}(M) \oplus \text{u}(1) \oplus ( N , \bar{M} ) \oplus
 ( \bar {N} , M ) \, .
 \label{suMdec}
\end{align}
We use $\alpha = 1 ,2 ,\ldots , N^2 -1 $ for the adjoint representation of $\text{su}(N)$
and $a , \bar a = 1, 2, \ldots , N$ for (anti-)fundamental representation of $\text{su}(N)$.
In addition to these $\text{su}(N)$ indices, we introduce $\text{su}(M)$ indices.
We use $\rho = 1 ,2 ,\ldots , M^2 -1 $ for the adjoint representation of $\text{su}(M)$
and $i , \bar \imath = 1, 2, \ldots , M$ for (anti-)fundamental representation of $\text{su}(M)$.
In total, the generators su$(N+M)$ can then be written as $\{t^\alpha , t^\rho , t^{\text{u}(1)}, t^{(a \bar \imath )} , t^{(\bar a i)}\}$ where the normalization is given in \eqref{suMgen}.

We introduce free fermions $\Psi^A$ in the adjoint representation of $\text{su}(N+M)$, i.e. $A=1,2,\ldots , (N+M)^2 -1$. The operator products are given by
\begin{align}
 \Psi^A (z) \Psi^B (w) \sim \frac{\delta^{AB}}{z-w} \, .
\end{align}
Then at level $k=N+M$ the $\text{su}(N+M)$  currents can be expressed by the free fermions as
\begin{align}
  J^A = - \frac{i}{2} \sum_{B,C} f^{ABC} \Psi^B \Psi^C \, .
\end{align}
The fermions are transforming in the adjoint under these currents
\begin{align}
 J^A (z) \Psi^B (w) \sim \sum_{C} i f^{ABC} \Psi^C (w) \frac{1}{z-w} \, .
\end{align}
The operator products between the currents can be computed as
\begin{align}
  J^A (z) J^B (w)
\sim
  \frac{(N+M) \delta^{AB}}{(z-w)^2}
    + \frac{\sum_C i f^{ABC} J^C (w)}{z-w} \, ,
\end{align}
which are indeed those for  $\text{su}(N+M)$ currents with level $k=N+M$.
{}For the computation, it is useful to utilize the formulas \eqref{formula1} and \eqref{formula2}.

The $\text{su}(N+M)_{N+M}$ currents can be decomposed according to \eqref{suMdec} as
\begin{align} \label{cosetMcurrents}
& J^\alpha = J^\alpha_1 + J^\alpha_2 \, , \quad
 J^\alpha_1 = - \frac{i}{2}  \sum_{\beta , \gamma} f^{\alpha \beta \gamma}
 \Psi^\beta \Psi^\gamma \, , \quad
 J^\alpha_2 = \Psi^{(b \bar \imath )} t^\alpha_{b \bar a} \Psi^{( \bar a j ) } \delta_{\bar \imath j}  \, , \\
& J^\rho = J^\rho_1 + J^\rho_2 \, , \quad
 J^\rho_1 = - \frac{i}{2}  \sum_{\sigma , \delta} f^{\rho \sigma \delta }
 \Psi^\sigma \Psi^\delta \, , \quad
 J^\rho_2 = \delta_{\bar a  b} \Psi^{(\bar a  i )} t^\rho_{i \bar \jmath} \Psi^{(b \bar \jmath ) }  \, , \nonumber
 \end{align}
 where $J^\alpha_1$ and $J^\alpha_2$ can be thought as the free fermion realization of su$(N)$ currents with level $N$ and $M$, respectively.
Similarly, $J^\rho_1$ and $J^\rho_2$ can be identified as su$(M)$ currents with level $M$ and $N$, respectively.
The other components of $\text{su}(N+M)_{N+M}$ currents are
\begin{align}
& J^{\text{u}(1)} = \sqrt{\frac{N+M}{NM}}  \Psi^{(a \bar \imath )} \Psi^{( \bar b j )} \delta_{a \bar b} \delta_{\bar \imath j} \, ,
\label{cosetMcurrents2} \\
& J^{(a \bar \imath ) } = \sqrt{\frac{N+N}{NM}} \Psi^{\text{u}(1)} \Psi^{ (a \bar \imath )} + \sum_\alpha
 \Psi^\alpha \Psi^{(b \bar \imath ) } t^\alpha_{b \bar a} \delta^{\bar a a} - \sum_\rho \Psi^\rho \delta^{\bar \imath i} t^\rho_{i \bar \jmath}\Psi^{(a \bar \jmath )}\, , \nonumber \\
& J^{(\bar a i )} = - \sqrt{\frac{N+N}{NM}} \Psi^{\text{u}(1)} \Psi^{ (\bar a i ) }  - \sum_\alpha
\delta^{\bar a a} \Psi^\alpha t^\alpha_{a \bar b} \Psi^{(\bar b i )} +   \sum_\rho \Psi^\rho \Psi^{(\bar a j )} t^\rho_{j \bar \imath} \delta^{\bar \imath i} \, . \nonumber
\end{align}
The energy momentum tensor is given by the coset construction of Sugawara Virasoro tensors
\begin{align}
   \label{cosetMEM}
 T &= T^{\text{su}(N+M)}_{N+M} - T^{\text{su}(N)}_{N+M} - T^{\text{u}(1)}_{NM(N+M)^2} \\
  &= \frac{1}{4(N+M)} \sum_A J^A J^A - \frac{1}{2(2N+M) } \sum_\alpha J^\alpha J^\alpha
   - \frac{1}{2NM(N+M)^2} \tilde J^{\text{u}(1)} \tilde J^{\text{u}(1)}  \nonumber
\end{align}
with
\begin{align}
 \tilde J^\text{u(1)} = \sqrt{NM(N+M)} J^{\text{u}(1)} \, .
\label{cosetMu1}
\end{align}
Since now we express the currents by free fermions, this energy momentum tensor can be
written only in terms of free fermions utilizing \eqref{cosetMcurrents} and
\eqref{cosetMcurrents2}.

We can also go back from the energy momentum tensor expressed by free fermions
to one by currents, but it does not always give the original expression.
With the formula
 (see, e,g, \cite{cft}  and (6.34) of \cite{Creutzig:2013tja})
\begin{align}
\label{peter}
 &  \frac{\delta_{a \bar b} \delta_{\bar \imath j}}{2} \left[ \partial \Psi^{ (a \bar \imath )}  \Psi^{ (\bar b j) }
    -  \Psi^{ (a \bar \imath )} \partial \Psi^{ (\bar b j) } \right] \\
  &  \qquad  = \frac{1}{2(N+M)}\sum_\alpha  J^\alpha_2 J^\alpha _2 + \frac{1}{2(M+N)} \sum_\rho J^\rho_2 J^\rho_2
   +\frac{1}{2NM(N+M)^2} \tilde J^{\text{u}(1)} \tilde J^{\text{u}(1)} \, ,
 \nonumber
\end{align}
we rewrite the energy momentum tensor \eqref{cosetMEM} of the model as
\begin{align}
 T &= T_D + T^{\text{su}(M)}_{N} +\sum_\rho T_{\Psi^\rho}+T_{\Psi^{\text{u}(1)}}  \, .\label{emad}
\end{align}
Here $T_D$ denotes the energy momentum tensor of \eqref{dcoset} with $k=N$ in \eqref{AhnEM},
where we have identified the su$(N)_M$ currents as $K^\alpha =J_2^\alpha$.
The energy momentum tensor
$T^{\text{su}(M)}_{N} $ is given by the Sugawara construction from the $\text{su}(M)_{N} $
currents $J^\rho_2$.  The last two are those for free fermions as
\begin{align}
 T_{\Psi^\rho} =  - \frac12  \Psi^{\rho} \partial \Psi^{\rho}  \, , \quad
 T_{\Psi^\text{u}(1)} =  - \frac12  \Psi^{\text{u}(1)} \partial \Psi^{\text{u}(1)} \, .
\end{align}
In summary, we have seen that $\text{su}(N+M)_{N+M}$, that is $(N+M)^2-1$ free fermions in the adjoint representation of $\text{SU}(N+M)$, decompose as follows: $N^2-1$
free fermions in the adjoint representation of $\text{SU}(N)$ forming $\text{su}(N)_N$; $M^2-1$
free fermions $\Psi_\rho$ in the adjoint representation of $\text{SU}(M)$ forming $\text{su}(M)_M$; $NM$ fermions in the tensor product of the fundamental representation of $\text{SU}(N)$ and the anti-fundamental representation of $\text{SU}(M)$ and another $NM$ fermions in its conjugate representation and finally one fermion $\Psi^{\text{u}(1)}$ in the trivial representation of both $\text{SU}(N)$ and $\text{SU}(M)$. So that the coset splits as
\begin{align}
\frac{\text{su}(N+M)_{N+M}}{\text{su}(N)_{N+M} \oplus \text{u}(1)_{NM(N+M)^2}}
 \approx \left[
 \frac{\text{su}(N)_{N} \oplus \text{su}(N)_M}{\text{su}(N)_{N+M}} \oplus \text{su}(M)_N \right] \oplus \Psi^{\rho} \oplus \Psi^{\text{u}(1)}.
 \label{proposal0}
\end{align}
In the right hand side, su$(N)_N$ and su$(N)_M$ are generated by $J_1^\alpha$ and $J_2^\alpha$
and the su$(N)_{N+M}$ is generated by $J^\alpha =J_1^\alpha + J_2^\alpha$ which also appears
in the left hand side. We know that the coset \eqref{dcoset} with $k=N$
is dual to \eqref{gcoset} with $k=N$ which has su$(M)_N$ in the denominator.
Let us assume that the su$(M)_N$ is canceled with the su$(M)_N$ in the right hand side of
\eqref{proposal0} as
\begin{align}
 \frac{\text{su}(N)_{N} \oplus \text{su}(N)_M}{\text{su}(N)_{N+M}} \oplus \text{su}(M)_N
 \approx
 \frac{\text{su}(N+M)_{N}}{\text{su}(N)_{N} \oplus \text{u}(1)_{N^2M(N+M)}} \, .
 \label{proposal1}
\end{align}
The su$(N)_M$ currents $J_2^\alpha$ are constructed by fermions in the bi-fundamental representation
 under the su$(M)$ invariant condition.  The su$(M)_N$ currents $J_2^\rho$ are also
constructed  by the same bi-fundamental fermions but with su$(N)$ invariant condition,
so we may argue that the currents play a role to relax the su$(M)$ invariant condition.
Once we admit this assumption, we arrive at \eqref{eq:compboscoset} using \eqref{proposal0}
and \eqref{proposal1}, where the decoupled fermions are $\Psi^\rho$ and $\Psi^{\text{u}(1)}$. Since ${\cal H}_{\Lambda_N}$ in \eqref{bspe} can be generated by the action of adjoint fermions $\Psi^A$ to the state $(0;\Lambda_N)$,%
\footnote{The label $\Lambda_N$ for $\text{su}(N)_{N+M}$ should be related to the label $\Lambda_{N+M}$ in \eqref{bspectrum}  for $\text{su}(N+M)_N$ by the action of transpose as usual for the level-rank duality, see, e.g., \cite{Naculich:1997ic}.}
 the Hilbert space after decoupling 
the fermions is \eqref{bspe} but with ${\cal H}_{\Lambda_N}$ 
generated by the action of $\Psi^\alpha$, $\Psi^{(a \bar \imath)}$ and $\Psi^{(\bar a i)}$.

\subsubsection{Generators at low spins}

Based on these preparations, particularly the relation \eqref{eq:compboscoset},
we can construct symmetry generators of the coset \eqref{bcoset} with $k = N$
from the one with $k=N+M$.
The symmetry generators are then constructed as combinations of the free fermions
in the adjoint representation of $\text{su}(N+M)$ that have only regular OPEs with
the $\text{su}(N)$ currents $J^\alpha$ in \eqref{cosetMcurrents} and the $\text{u}(1)$ current $J^\text{u(1)}$ in \eqref{cosetMcurrents2}. Furthermore, we need to decouple free fermions
$\Psi^\rho$ and $\Psi^{\text{u}(1)}$ according to \eqref{eq:compboscoset}.

Let us start from spin $1/2$ generators. {}From the coset with $k=N+M$,
we find that the spin $1/2$ generators are given by
\begin{align}
\Psi^\rho \, , \quad \Psi^{\text{u}(1)} \, .
\end{align}
They are exactly the fermions that we decouple, so there are no spin $1/2$ generators.
Since the dual gravity theory does not have any spin 1/2 gauge fields,
it is consistent with the proposed duality.

We then move to spin $1$ generators.
Spin $1$ currents can be constructed by bilinears of fermions.
We find that they are  given by $J_2^\rho$ in \eqref{cosetMcurrents},
\begin{align}
 J_2^\rho = \delta_{\bar a b} \Psi^{(\bar a i )}  t^\rho_{i \bar \jmath}  \Psi^{(b \bar \jmath)} \, ,
\end{align}
which generates the $\text{su}(M)$ current algebra with level $N$.
They are expected to be dual to the $M^2 -1$ spin 1 gauge fields in the dual gravity theory.

Spin $3/2$ supercurrents consist of products of three fermions.
We find the following fields
\begin{align}
  G^{i \bar \jmath} =
 \delta^{i \bar \jmath} \Psi^\alpha J_1^\alpha + 3 N  t^\alpha_{b \bar a}  \Psi^\alpha \Psi^{(\bar a i )} \Psi^{( b \bar \jmath )}  \,  , \label{scg1}
\end{align}
where the relative coefficient is fixed so as to have only regular OPEs with the su$(N)$ and u$(1)$ currents,
see also appendix \ref{App:Ahn}.
They are expected to be dual to the $M^2$ spin $3/2$ gauge fields in the higher spin supergravity theory.

As we will see in section \ref{gravity2}, we pick up so$(2n+1)$ generators, such as $Q^I_\alpha$ in \eqref{ospi} among the u$(M)$ with $M=2^n$ supercurrents as supersymmetry generators in the gravity side.
With the restricted class of spin 3/2 generators, we have ${\cal N} =2n+1 $ supersymmetry algebra
as a consistent truncation of the whole higher spin symmetry algebra.
However, there seems to be no such a consistent truncation in the bosonic coset \eqref{bcoset} with
finite $k=N,M$ because of the non-linear terms in the symmetry algebra.

Among the linear combinations of the spin 3/2 generators in \eqref{scg1}, there is a special operator defined by
\begin{align}
 G = \frac{\tilde C \delta_{i \bar \jmath} }{M} G^{i \bar \jmath}
  =  \tilde C \sum_\alpha \Psi^\alpha \left(  J_1^\alpha - \frac{3N}{M} J_2^\alpha \right) \label{cosetMG}
\end{align}
with $\tilde C$ defined in \eqref{scgd}.
Due to the identification $J_2^\alpha = K^\alpha$, we have $G =G_D$ in
\eqref{scgd}.
As in the relation \eqref{proposal1}, the bosonic coset \eqref{bcoset}  can be decomposed into
a coset model and su$(M)$ currents. The coset model is known to have ${\cal N}=1$ superconformal
 symmetry \cite{Goddard:1986ee,Douglas:1987cv} and \eqref{cosetMG} is identified as the superconformal generator \cite{Ahn:2012bp,Ahn:2013ota}. For $M=1$, there is no su$(M)$ sector, and this is the ${\cal N}=1$ superconformal
 symmetry used in \cite{Beccaria:2013wqa}. If we forget about spin 1 gauge fields, then there is ${\cal N }=1$ supersymmetry between spin $s$ and spin $s-1/2$ gauge fields even with $M \neq 1$.
 Thus the ${\cal N}=1$ supersymmetry structure in the bosonic coset \eqref{bcoset} maps nicely that in the gravity theory. Note, however, that the  ${\cal N}=1$ supersymmetry is generated by $Q^{{\cal N}=1}_\alpha = y_\alpha \otimes \1$
 in terms of \eqref{yphi}, and it is not a part of ${\cal N} =2n+1 $ supersymmetry generated by $Q^I_\alpha$ in \eqref{ospi}. In other words, if we include both the ${\cal N}=1$ generator $Q^{{\cal N}=1}_\alpha$ and the ${\cal N}=2n+1$ generators $Q_\alpha^I$, then we do not have any supersymmetry algebra as a
 consistent truncation up to spin 2 generators.

We can construct more higher spin operators in a similar way.
In particular, one of the spin 2 operator
is given in terms of Sugawara energy-momentum tensor as in \eqref{cosetMEM}.

Before ending this section, we would like to summarize the results so far.
We have considered two coset models
\begin{itemize}
\item[(B1)] The bosonic coset \eqref{bcoset} with $k=N$ and the spectrum in \eqref{bspectrum}.
\item[(B2)] The bosonic coset \eqref{bcoset} with $k=N+M$, but several fermions decoupled.
\end{itemize}
Our conjecture is that the higher spin gravity with ${\cal N}=2n+1$ is dual to the model (B1). This candidate is a natural extension of the $n=0$ case in \cite{Beccaria:2013wqa}, and support for this conjecture is given by the match of one-loop partition function as shown in subsection \ref{bpf}. The model (B1) is not suitable for examining its enhanced symmetry, so we utilize the other model (B2) for this purpose. The claim here is that the models (B1) and (B2) are really the same, see \eqref{eq:compboscoset}, namely they have the same spectrum, symmetry and so on, even for finite $N$. However, we should say that the derivation is not well supported since we have heavily relied on the bosonic level-rank duality in \cite{Bowcock:1988vs,Altschuler:1988mg} which is not well understood yet.
In the next section we use the supersymmetric coset of the form \eqref{coset} instead of \eqref{bcoset}. We do this because the level-rank duality is well understood in the supersymmetric case as in \cite{Kazama:1988qp,Naculich:1997ic}. Moreover, we expect that it is easier to see the relation to superstring theory with the supersymmetric coset.

\section{The supersymmetric coset models}
\label{scm}

In this section we examine the supersymmetric coset \eqref{coset} at special levels $k=N\pm M$,
and the Kazama-Suzuki model \eqref{KScoset} with $k=N+M$.
In the next subsection, we compute the partition function of the coset \eqref{coset} with $k=N-M$ and large $N$,
and show that the result matches with the gravity partition function in \eqref{eq:partrewritten} with $M$ replaced by
$2M$  and with the contribution from some free fermions removed.
In subsection \ref{sec:ssg}, we first obtain the relation between the cosets \eqref{coset} with $k=N-M$
and $k=N+M$ by utilizing a level-rank duality in \cite{Kazama:1988qp,Naculich:1997ic}.
Then we show that these models with some free fermions decoupled
are related to the bosonic model \eqref{bcoset} with $k=N$ and $M$ replaced by $2M$,
see \eqref{eq:compsupercoset}. With these relations and results for the bosonic coset,
we can obtain symmetry generators of the supersymmetric cosets \eqref{coset} at the specific levels.
In subsection \ref{sec:N=3}, we show that the Kazama-Suzuki model \eqref{KScoset} with $k=N+M$ has ${\cal N}=3$ superconformal symmetry even at finite $N$.
Based on this finding, we conjecture that the critical level model is dual to a superstring theory, and examine the
duality by comparing BPS states and marginal deformations in subsection \ref{sec:superstrings}.

\subsection{Spectrum and partition functions}

We examine the spectrum and the one-loop partition function of the supersymmetric coset \eqref{coset}.
We take the large $N$ limit with $k=N-M$ but keep $M$ finite.
As before we denote the highest weights of su$(L)$ by $\Lambda_L$ and
$m \in \mathbb{Z}_\kappa$. Moreover, by NS we denote the sum of
identity and vector representations of so$(2NM)_1$, which are generated by
bi-fundamental free fermions $\psi^{(a \bar \imath )}, \psi^{ ( \bar a i ) }$.
Then the states of
the coset are given by the decomposition
\begin{align}
 \Lambda_{N+M} \otimes \text{NS} =
\bigoplus_{\Lambda_N ,m } (\Lambda_{N+M};\Lambda_N ,m) \otimes \Lambda_{N+M} \otimes m \, .
\end{align}
At large $N$, $m$ is fixed as in \eqref{mfix}, thus
the states are labeled by $(\Lambda_{N+M};\Lambda_N)$ in the large $N$ limit.
As in the bosonic case, we consider the following spectrum as
\begin{align}
 {\cal H} = \bigoplus_{\Lambda_{N+M}}  {\cal H}_{\Lambda_{N+M}} \otimes \bar {\cal H}_{\Lambda_{N+M}^*}
\, , \quad
 {\cal H}_{\Lambda_{N+M}} = \bigoplus_{\Lambda_N \in \Omega} (\Lambda_{N+M};\Lambda_N) \, ,
\label{spectrum}
\end{align}
where $\Omega$ means that we take the sum over $\Lambda^l_N = (\Lambda^r_N)^t$.

In the 't Hooft limit with \eqref{thooft}, the character of $(\Lambda ; \Xi)$ can be computed as \cite{Creutzig:2013tja}
\begin{align}
 sb^{\lambda,M}_{\Lambda ; \Xi} (q) = {\cal X}^M_0(q) q^{\frac{\lambda}{2} (|\Lambda^l| + |\Lambda^r|-|\Xi^l| - |\Xi^r| )}
\sum_{\Phi , \Psi} R^{(N)}_{\Lambda \Phi} C^{(N)\Psi^*}_{\Phi \Xi^*} {\cal B}^{1/2,M}_{(\Psi^l)^t} (q) {\cal B}^{1/2,M}_{(\Psi^r)^t} (q) \, ,
\end{align}
where $R^{(N)}_{\Lambda \Phi}$ and $C^{(N)\Psi^*}_{\Phi \Xi^*}$ are introduced in
\eqref{restriction} and ${\cal B}^{h,M}_{\alpha} (q)$ is defined in \eqref{sBhM}.
The vacuum character is
\begin{align}
 sb^{\lambda,M}_{0; 0} (q) =   {\cal X}^M_0 (q) = \left(\prod_{s=2}^\infty z_{B}^{(s)} (q)  z_F^{(s)} (q)\right )^{2M^2 }
 \left(z_{B}^{(1)}(q)\right)^{2M^2 - 1} \, .
\end{align}
{}From the spectrum we consider
\begin{align}
 s\chi^M_\Lambda (q) = \sum_{\Xi \in \Omega} sb^{1/2,M}_{\Lambda ; \Xi} (q) \, .
 \label{schiM0}
\end{align}
As shown in appendix \ref{PF}, this expression can be reduced to
\begin{align}
 s \chi_\Lambda^M (q) = \chi_0^{2M} (q) ( z_F^{(1)} (q) )^{M^2}
{\cal B}^{1/4 ,2M}_{\Lambda^l} (q) {\cal B}^{1/4 ,2M}_{\Lambda^r} (q) \, .
\label{schiM}
\end{align}
Here $( z_F^{(1)} (q) )^{M^2}$ is the contribution from  $M^2$ spin $1/2$ fermions,
and it is removed as we will discuss below.
The large $N$ limit of  one-loop partition function of the coset \eqref{coset} with the
spectrum \eqref{spectrum} and with fermions decoupled is thus
\begin{align}
s{\cal Z}_\text{CFT} (q) = \left| \chi^{2M}_0 (q) \right| ^2 \sum_{\Lambda^l , \Lambda^r} \left|  {\cal B}^{1/4,2M}_{\Lambda^l} (q)
{\cal B}^{1/4,2M}_{\Lambda^r} (q) \right|^2 \, .
\label{sCFTPF}
\end{align}
This reproduces the one-loop partition function \eqref{eq:partrewritten} for ${\cal N}=2n+1$ higher spin
supergravity where $n$ is determined by $M = 2^{n-1}$.
Let us stress here that the supersymmetric coset \eqref{coset} with parameter $M$ is mapped to
higher spin supergravity with U$(2M)$ Chan-Paton factor instead of U$(M)$ as in the bosonic case.

\subsection{Symmetry generators}
\label{sec:ssg}

In the previous subsection, we have computed the partition function for $k=N-M$. However, in that case the critical level factor appears in the denominator. For the analysis of the symmetries of the coset,
we find a map from the coset into the one with the critical level $k=N+M$ of su$(N+M)_k$
in the numerator.

We start from the ${\cal N}=2$
Grassmannian Kazama-Suzuki coset \eqref{KScoset}.
The coset \eqref{coset} is simply given by adding the $\text{su}(M)_{k+N}$ factor.
Here  we show that the relation between the two conditions $k=N\pm M$ simply is given by a translation $N\mapsto N+M$.
To see this, we remember that the coset \eqref{KScoset} is level-rank dual to \cite{Kazama:1988qp,Naculich:1997ic}
\begin{align}
 \frac{\text{su}(k+M)_N \oplus \text{so}(2kM)_1}{\text{su}(k)_{N+M} \oplus
\text{su}(M)_{N+k} \oplus \text{u}(1)_\kappa} \, ,
\label{dKScoset}
\end{align}
which is obtained by replacing $N$ and $k$. If we use the condition $k=N-M$ in this expression and set
$N = \tilde N + M$, we have
\begin{align}
 \frac{\text{su}(\tilde N+M)_{\tilde N + M} \oplus \text{so}(2\tilde NM)_1}{\text{su}(\tilde N)_{\tilde N + 2 M} \oplus
\text{su}(M)_{M + 2 \tilde N} \oplus \text{u}(1)_\kappa}
\label{dKScoset2}
\end{align}
which is the same as \eqref{KScoset} with $N = \tilde N$ and $k = \tilde N + M$.
Notice that the same factor $\text{su}(M)_{2N-M} = \text{su}(M)_{2 \tilde N + M}$ appears
both in \eqref{dKScoset} and \eqref{dKScoset2}.
Therefore, adding the same factor  $\text{su}(M)_{2N-M}  = \text{su}(M)_{M + 2 \tilde N}$ we have a dual coset description as
\begin{align}
 \frac{\text{su}(\tilde N+M)_{\tilde N + M} \oplus \text{so}(2\tilde NM)_1}{\text{su}(\tilde N)_{\tilde N + 2 M} \oplus
 \text{u}(1)_\kappa} \, ,
\end{align}
which is the supersymmetric coset \eqref{coset} with $k=\tilde N +M$ and $N = \tilde N$.
Since the factor described by adjoint free fermions then appears in the numerator of the coset,
it should not be so difficult to construct generators of symmetry algebra.
Indeed we can see this by relating to the analysis of the bosonic case in the previous section, as done below.

In \eqref{proposal1}, we assumed that the $\text{su}(M)_N$ currents in the left hand side cancel
the $\text{su}(M)_N$ in the denominator of \eqref{gcoset} after using the level-rank duality
with \eqref{dcoset}.
However the factor $\text{su}(M)_k$ in the denominator of \eqref{gcoset} does not appear in \eqref{dcoset}.
Therefore, we may say that the cancellation of the su$(M)$ factor after applying the level-rank duality
becomes more transparent in these supersymmetric coset models.

At the level $k=N+M$, the $\text{su}(N+M)_k $ currents can be described by free fermions
\begin{align}
 \Psi^\alpha \, , \quad \Psi^{(a \bar \imath)} \, , \quad \Psi^{(\bar a  i)} \, , \quad \Psi^\rho \, , \quad \Psi^{\text{u}(1)} \, .
\end{align}
 Since $\text{so}(2NM)_1$ can be described by bi-fundamental free fermions $\psi^{ ( a \bar \imath ) }, \psi^{ ( \bar a i )}$,
we have in total $4NM$ bi-fundamental fermions as
\begin{align}
 \Psi^{(a \bar \imath)} \, , \quad \Psi^{(\bar a  i)} \, , \quad
  \Psi^{(a  (M + \bar \imath ))} =  \psi ^{(a \bar \imath)}  \, , \quad \Psi^{(\bar a  (M + i )) } = \psi^{(\bar a i)}
  \label{Psinotation}\, .
\end{align}
We can easily see that the coset is the same as the bosonic coset \eqref{bcoset} with $M$ replaced by $2M$ after decoupling
free fermions $ \Psi^\rho \, , \Psi^{\text{u}(1)} $ (up to a normalization of the $\text{u}(1)$ current).
Extending equation \eqref{eq:compboscoset}, we thus have the relation
\begin{align}
   \frac{\text{su}(N+2M)_N}{\text{su}(N)_{N} \oplus \text{u}(1)_{N^22M(N+2M)}}&\approx \frac{\text{su}(N+2M)_{N+2M}}{\text{su}(N)_{N+2M} \oplus \text{u}(1)_{N2M(N+2M)^2}}\Big|_{\text{Free fermions decoupled}}\nonumber\\
   &\approx \frac{\text{su}(N+M)_{N+M} \oplus \text{so}(2NM)_1}{\text{su}(N)_{N+2M} \oplus \text{u}(1)_\kappa}\Big|_{\text{Free fermions decoupled}}  \label{eq:compsupercoset}
\end{align}
Thus the analysis from the previous section actually also applies here by just changing
the interpretation of half of the bi-fundamental fermions.
For instance we can find out the symmetry generators of the coset \eqref{coset} in this way. Moreover, the Hilbert space should be  ${\cal H} = \bigoplus_{\Lambda_{N}}  {\cal H}_{\Lambda_{N}}
\otimes \bar {\cal H}_{\Lambda_{N}^*}$ with ${\cal H}_{\Lambda_N}$ generated by the action of fermions other than $ \Psi^\rho \, , \Psi^{\text{u}(1)} $ to the state $(0;\Lambda_N)$.

When we relate the model to the bosonic coset \eqref{bcoset} with $k=N$,
we decoupled $M^2$ free fermions, $ \Psi^\rho \, , \Psi^{\text{u}(1)} $.
The contribution from these free fermions to \eqref{schiM} is removed in order to obtain
\eqref{sCFTPF}, which reproduces the gravity partition function.
This is consistent since the bosonic coset \eqref{bcoset} with $k=N$
is already shown to be dual to the higher spin gravity in the previous section.

In addition to the two models (B1) and (B2) in the end of the previous section, we now have two additional models as
\begin{itemize}
\item[(S1)] The supersymemtric coset \eqref{coset} with $k=N-M$ and the spectrum \eqref{spectrum}. Moreover, some fermions are decoupled.
\item[(S2)] The supersymmetric coset \eqref{coset} with $k=N+M$ with even more fermions decoupled.
\end{itemize}
We propose that the model (S1) is another candidate for the CFT dual of the higher spin theory with ${\cal N}=2n+1$ supersymmetry along with the model (B1). The model (S2) is again used for the study of enhanced symmetry, and it will be utilized to see the relation with superstring theory as below. Our claim is that the models (S1) and (S2) are equivalent. Since the level-rank duality in \cite{Kazama:1988qp,Naculich:1997ic} is better understood in the supersymmetric case, we think the equivalence is more reliable than the relation between the models (B1) and (B2). It is easy to relate (S2) to (B2) just by changing the interpretation of half of the free fermions, so we may be able to say that the models (S1) and (B1) are actually the same by applying the chain of relations.

\subsection{${\cal N}=3$ enhanced supersymmetry}
\label{sec:N=3}

One of the aim of this paper is to find out a triality between 3d Vasiliev theory, superstring theory
and a 2d conformal model just like the ABJ triality in \cite{Chang:2012kt}.
We have proposed that the 3d Vasiliev theory introduced in section \ref{gravity1} is dual to
the supersymmetric coset \eqref{coset} with $k=N+M$ and with some fermions decoupled
(or equally the bosonic coset model \eqref{bcoset} with $k=N$).
In the case of ABJ triality, only su$(M)$ singlet combinations of higher spin gauge fields
are dual to string states. Therefore, it is natural to think that a superstring theory is dual to
the su$(M)$ gauged version of the coset \eqref{coset}, that is the Kazama-Suzuki model \eqref{KScoset}.%
\footnote{In this and the next subsections, we consider generic integer $M$ since we compare the Kazama-Suzuki model \eqref{KScoset} not to higher spin theory, but to superstring theory.}

In order to construct the Grassmannian coset \eqref{KScoset}, we simply introduce
${\cal N}=1$ worldsheet supersymmetry. However due to the Kazama-Suzuki construction
\cite{Kazama:1988uz,Kazama:1988qp}, the coset actually has
${\cal N}=2$ enhanced superconformal symmetry.
In this subsection, we would like to show that the superconformal symmetry enhances from
${\cal N}=2$ to ${\cal N}=3$ for the Grassmannian Kazama-Suzuki model \eqref{KScoset}
with $k=N+M$ even with finite $N$ (and without decoupling fermions).
The ${\cal N}=3$ superconformal symmetry quite restricts
the target space of dual string theory,
which is of the form AdS$_3 \times $M$_7$.
In the next subsection we will examine the relation between the Kazama-Suzuki model \eqref{KScoset} with $k=N+M$ and superstring theory on AdS$_3 \times $M$_7$
by comparing BPS states and marginal deformations.

The Grassmannian Kazama-Suzuki coset \eqref{KScoset}
with $k=N+M$ has the central charge of the simple form
\begin{align}
 c = \frac{3}{2} NM \, .
\label{centerKS}
\end{align}
At this level the $\text{su}(N+M)_{N+M}$ currents in the numerator can be described by free fermions
in the adjoint representation of $\text{su}(N+M)$ as argued above.
Thus the symmetry generators are constructed by these free fermions
in addition to $\psi^{(a \bar \imath)}, \psi^{(\bar a i)}$
coming from $\text{so}(2NM)_1$.
Moreover, they should have only regular OPEs with currents in the denominator.
Defining three sets of currents as
\begin{align}
 j^\alpha = \psi^{(b \bar \imath )} t^\alpha_{b \bar a} \psi^{(\bar a j)} \delta_{\bar \imath j} \, , \quad
 j^\rho =  \delta_{\bar a b}\psi^{(\bar a i )}
  t^\rho_{i \bar \jmath}  \psi^{(b \bar \jmath)} \, , \quad
j^{\text{u}(1)} = \psi^{(a \bar \imath)} \psi^{(\bar b j)} \delta_{a \bar b} \delta_{\bar \imath j} \, ,
\end{align}
the $\text{su}(N)$, $\text{su}(M)$ and $\text{u}(1)$ currents in the denominator are expressed as
\begin{align}
  J^\alpha + j^\alpha \, , \quad   J^\rho + j^\rho \, , \quad
\tilde J^{\text{u}(1)} + (N+M) j^{\text{u}(1)}
 \equiv  (N+M) [ \hat J^{\text{u}(1)} + j^{\text{u}(1)} ] \, ,
 \label{denominator}
\end{align}
respectively. Here $J^\alpha$, $J^\rho$, and $\tilde J^{\text{u}(1)}$ are defined in
\eqref{cosetMcurrents} and \eqref{cosetMu1} and $\hat J^{\text{u}(1)}$
is defined in the above equation.

Let us construct symmetry generators explicitly.
Due to the Kazama-Suzuki construction  \cite{Kazama:1988uz,Kazama:1988qp},
the model at least has ${\cal N}=2$ superconformal symmetry generated by a spin 1 $R$-current $J$,
two spin 3/2 superconformal currents $G^\pm$, and a spin 2 energy momentum tensor $T$.
The spin 1 current is
\begin{align}
 J = \frac12 \delta_{a \bar b} \delta_{\bar \imath j} \left( \Psi^{(a \bar \imath )} \Psi^{( \bar b j )}
 -  \psi^{( a \bar \imath )} \psi^{( \bar b j )} \right )\, ,
 \label{N=2J}
\end{align}
which satisfies
\begin{align}
J (z) J(w) \sim \frac{c}{ 3 (z-w)^2}
\end{align}
with $c= 3 N M / 2$ as in \eqref{centerKS}.
The spin 3/2 currents are given by
\begin{align}
 G^+ = \frac{1}{\sqrt{N + M}} \delta_{a \bar b} \delta_{\bar \imath j}J^{ ( a \bar \imath )} \psi^{ ( \bar b j )} \, , \quad
 G^- =  \frac{1}{\sqrt{N + M}}\delta_{\bar a  b} \delta_{i \bar \jmath } J^{(\bar  a i )} \psi^{ ( b \bar \jmath ) } \,
\end{align}
where $J^{(a \bar \imath )} , J^{ ( \bar a i ) }$ are defined in \eqref{cosetMcurrents2}.
The spin 2 generator is written in terms of the Sugawara operators as
\begin{align}
\label{N=3Sugawara}
 T =  & \frac{1}{4(N+M)} \left[ \sum_\alpha J^\alpha J^\alpha + \sum_\rho J^\rho J^\rho +2 \delta_{a \bar b} \delta_{\bar \imath j}J^{(a \bar \imath)} J^{(\bar b j)}
+ J^{\text{u}(1)} J^{\text{u}(1)} \right]  \\
 & -  \frac12 \delta_{a \bar b}  \delta_{\bar \imath j} \left[   \psi^{(a \bar \imath)} \partial \psi^{(\bar b j)} -  \partial \psi^{(a \bar \imath)} \psi^{(\bar b j)} \right]
- \frac{1}{4(N+M)} \sum_\alpha (J^\alpha + j^\alpha ) ( J^\alpha + j^\alpha)  \nonumber \\
 & -  \frac{1}{4(N+M)} \sum_\rho (J^\rho + j^\rho ) ( J^\rho + j^\rho)
  - \frac{1}{4N M } (\hat J^{\text{u}(1)} + j^{\text{u}(1)})
(\hat J^{\text{u}(1)} + j^{\text{u}(1)})  \, .  \nonumber
\end{align}
In particular, we have the following OPE as
\begin{align}
G^+ (z) G^- (w) \sim \frac{2c/3}{(z-w)^3} + \frac{2J(w)}{(z-w)^2} +
\frac{2 T(w) + \partial J(w)}{z-w} \, .
\end{align}

The above currents generating ${\cal N}=2$ superconformal symmetry
exist even with generic level $k$. Here we would like to show that
there are additional currents at the specific level $k=N+M$, which are
given by a spin 1/2 fermion $\Psi$, two spin 1 currents $J^\pm$ and a spin 3/2 currents $G^3$.
Combined with the generators of the ${\cal N}=2$ superconformal symmetry,
they generate ${\cal N}=3$ superconformal symmetry.
The spin 1/2 generator is given by $\Psi = \Psi^{\text{u}(1)}$, and
the extra spin 1 currents are
\begin{align}
 J^+ =  \frac{1}{\sqrt2}\delta_{a \bar b} \delta_{\bar \imath j}  \Psi^{(a \bar \imath )} \psi^{ (\bar b j )} \, , \quad J^- =   \frac{1}{\sqrt2} \delta_{a \bar b} \delta_{\bar \imath j} \psi^{ ( a \bar \imath ) }  \Psi^{ ( \bar b j ) }
\end{align}
in addition to $J^3 = J$ in \eqref{N=2J}.
We find that last spin 3/2 generator  is given by
\begin{align}
  \label{G3}
 G^3 = \frac{1}{\sqrt{2(N + M)}}  \left[ \sum_\alpha \Psi^\alpha \left(J_2^\alpha
  -  j^\alpha \right)
   +  \sum_{\rho } \Psi^\rho \left(J_2^\rho
  -  j^\rho \right) \right]  \qquad \\
   + \frac{1}{\sqrt{2 N M}} \Psi^{\text{u}(1)}  \left( \hat J^{\text{u}(1)}
  - j^{\text{u}(1)} \right) \, . \nonumber
\end{align}
As shown in appendix \ref{sec:G3}, this superconformal generator satisfies
\begin{align}
G^3 (z) G^3 (w) \sim \frac{2c/3}{(z-w)^3}  +
\frac{2 T(w)}{z-w} \, .
\label{GGope}
\end{align}
Similarly we can show
that the generators given above also lead to the remaining OPEs of the $\mathcal{N}=3$ superalgebra given in appendix \ref{N=3} by setting
\begin{align}
 G^\pm =  \frac{1}{\sqrt2} ( G^1 \pm i G^2 ) \, , \quad J^\pm =  \frac{1}{\sqrt2} (J^1 \pm i J^2 )
\end{align}
and $c = 3 N M / 2$, $k=N M$.

The supercurrents of the coset \eqref{coset} are generically of the form as
\begin{align}
G[t^\rho] \sim t^\alpha_{b \bar a}  \Psi^{(i \bar a)} \Psi^{\alpha} \Psi^{(b \bar \jmath)} {(t_{2M})}^\rho_{i \bar \jmath}
\end{align}
whose precise expression is given in \eqref{scg1}. Here we use the notation  in \eqref{Psinotation} as
$
  \Psi^{(a  (M + \bar \imath ))} =  \psi ^{(a \bar \imath)}  , \Psi^{(\bar a  (M + i )) } = \psi^{(\bar a i)}
$,
and the u$(2M)$ generator $(t_{2M})^\rho$ corresponds to the Chan-Paton factor of higher spin
gravity theory.
In the model \eqref{KScoset}, we assign the su$(M)$ singlet condition.
Therefore only supercurrents with $(t_{2M})^\rho= \sigma^\delta \otimes
\1_M$ $(\delta=0,1,2,3)$ are allowed, where
$\sigma^a$ $(a=1,2,3)$ are the Pauli matrices
and $\sigma^0 = \1_2$. The ${\cal N}=3$ supercharges $G^a$  are associated with $\sigma^a$.
Notice that the ${\cal N}=1$ supercurrent $G^{{\cal N}=1}$ introduced in \eqref{cosetMG}
is with $\sigma^0 = \1_2$, and they are different from those of ${\cal N}=3$ supercurrents.

Each ${\cal N}=3$ superconformal current
\begin{align}
 G^1 =  \frac{1}{\sqrt2} ( G^+ + G^- ) \, , \quad
G^2 =  \frac{1}{\sqrt2 i}  (G^+ - G^-) \, , \quad G^3
\end{align}
generates ${\cal N}=1$ superconformal algebra as a subalgebra
\begin{align}
 G^a (z) G^a (w) \sim \frac{2c/3}{(z-w)^3} + \frac{2 T(w)}{z-w}
\end{align}
with $a$  not summed over.
Here $c$ and $T$ are the central charge in \eqref{centerKS} and the energy
momentum tensor in \eqref{N=3Sugawara} for the Kazama-Suzuki model \eqref{KScoset}
with $k=N+M$.
On the other hand, the ${\cal N}=1$ supercharge $G^{{\cal N}=1}$ satisfies the OPE \eqref{N=1OPE}
with $c_D$  in \eqref{centrald} and $T_\text{D}$ in \eqref{AhnEM} for the bosonic Grassmannian model \eqref{gcoset} or its dual form \eqref{dcoset}
with $k=N$. These two models \eqref{KScoset} and \eqref{gcoset} differ in particular by free fermions, so we can see also in this way that $G^a$ and $G^{{\cal N} =1}$
are essentially different operators at finite $N$.

One may think
that these four supercurrents would generate the small ${\cal N}=4$ superconformal
algebra. However, if we include both $G^a$ and $G^{{\cal N}=1}$, then the algebra does not seem to close up to spin 2 generators.

\subsection{Relation to superstring theory}
\label{sec:superstrings}
In this subsection, we study chiral primaries of the coset and compare them to BPS states from the dual
string theory.%
\footnote{We assume that both $N$ and $M$ are very large in this subsection.}
The states of the coset are labeled by $(\Lambda_{N+M}, \omega ; \Lambda_N , \Lambda_M , m)$, where $\Lambda_L$ denotes a highest weight of su$(L)$ as before. Moreover, we take $\omega = 0,2$
for the NS sector and $m \in \mathbb{Z}_{\kappa}$.
The conformal weight for the state with  $(\Lambda_{N+M}, \omega ; \Lambda_N , \Lambda_M , m)$ is
\begin{align}
 h = n + h^{N+N, N+M}_{\Lambda_{N+M}} + \frac{\omega}{4}
 - h^{N , N+2M}_{\Lambda_N} - h^{M , 2N+M}_{\Lambda_N}
 - h_m \, , \label{KScf1}
\end{align}
where
\begin{align}
 h^{L,K}_{\Lambda_L} = \frac{C^L(\Lambda_L)}{K+L} \, , \quad
 h_m = \frac{m^2}{2 \kappa} \, . \label{KScf2}
\end{align}
Here $C^L (\Lambda_L)$ is the quadratic Casimir of the representation $\Lambda_L$.
We do not have a general formula for integer $n$, but it is easy to compute for a specific
example by considering how the denominator is embedded in the numerator, see, for instance \cite{cft}.

We have introduced fermions $\Psi^{(a  \bar \imath)}, \Psi^{(\bar a i)}$ for su$(N+M)_{N+M}$ and  $\psi^{(a  \bar \imath)}, \psi^{(\bar a i)}$ for so$(2NM)_1$.
Moreover, $R$-current $J^3$ is written in terms of these fermions as $J^3 = J$ in \eqref{N=2J},
and we denote its eigenvalue  as $q$.
Here we look for chiral primaries which satisfies the BPS condition
$h = q/2$.
See \cite{Miki:1989ri} for the representation theory of ${\cal N}=3$ superconformal algebra.
We assume the form of chiral primary as $( \Psi^{(a  \bar \imath)})^p ( \psi^{(\bar a i)} )^l | v \rangle$, whose $R$-charge  is given by $q = (p+l)/2$.
As before, we  use the expression of $\Lambda_L$ by two Young diagrams
as $(\Lambda_L^l , \Lambda_L^r)$.
With the notation
\begin{align}
[p;l] \equiv ([p,0,\ldots , 0] , [ l , 0 , \ldots , 0]) \, , \quad
(l;p) \equiv ([ 0^{l-1} , 1 , 0 , \ldots , 0] , [ 0^{p-1} , 1 , 0 , \ldots , 0]) \, ,
\end{align}
we set $(\Lambda_N , \Lambda_M , m) = ([p;l], (l;p), (N+M)(l-p))$,
where we can check that the selection rules are satisfied, see \cite{Lerche:1989uy}.
More generic states will be mentioned later.
Using the formulas
\begin{align}
& C^{L} ([p,0,\ldots , 0]) = \frac{p(L-1)(L+p)}{2L} \, , \quad
 C^{L} ([ 0^{p-1} , 1 , 0 , \ldots , 0]) = \frac{p(L+1)(L-p)}{2L} \, ,
\end{align}
and
\begin{align}
& C^{L} (\Lambda_L) = C^{L} (\Lambda_L^l) + C^{L} (\Lambda_L^r)
 + \frac{|\Lambda_L^l| |\Lambda_L^r|}{L} \, ,
\end{align}
we can compute
the conformal weight as
\begin{align}
 h = \frac{p + l}{2} - \frac{C^N ([p;l]) }{2 (N+M)}  - \frac{C^M ((l;p)) }{2 (N+M)} - \frac{(N+M)^2 (l-p)^2 }{4 NM (N+M)^2}
=
\frac{p + l}{4} = \frac{q}{2} \, ,
\label{BPS}
\end{align}
which means that the corresponding states are chiral primaries.

Let us interpret these states in terms of dual gravity theories.
The simplest ones are with $(p,l) = (1,0)$ and $(0,1)$, which may be called as $|c_\eta \rangle$ with $\eta = 0,1$.
Similarly we have two simplest anti-chiral primaries $|a_\eta \rangle$.%
\footnote{Anti-chiral primaries are of the form as
$( \psi^{(a  \bar \imath)})^p ( \Psi^{(\bar a i)} )^l | v \rangle$. We should pair the state in the holomorphic sector with that in the anti-holomorphic sector labeled by the same $v$.
}
Combining the anti-holomorphic sector, we have  eight states
\begin{align*}
 | c_\eta \rangle \otimes | \bar c_\eta \rangle \, , \quad
 | c_\eta \rangle  \otimes | \bar a_\eta \rangle \, , \quad
 | a_\eta \rangle  \otimes | \bar c_\eta \rangle \, , \quad
 | a_\eta \rangle  \otimes | \bar a_\eta \rangle \, .
\end{align*}
We would like to propose that their duals
are given by scalar fields with dual conformal weight $(h,h)=(1/4,1/4)$ in the higher spin gravity introduced in section \ref{gravity1}.
For the dual of the coset \eqref{KScoset}, we have a $(2 M \times 2M)$ matrix
valued complex scalar field with a su$(M)$ invariant condition.
Therefore we have four (or eight in the real counting) su$(M)$ invariant combinations and two of them are
charged under the corresponding $R$-currents.
Thus we can identify the uncharged combinations as $ | c_\eta \rangle  \otimes | \bar a_\eta \rangle$,
$ | a_\eta \rangle  \otimes | \bar c_\eta \rangle$ and the charged ones as
$ | c_\eta \rangle \otimes | \bar c_\eta \rangle $ and  $| a_\eta \rangle  \otimes | \bar a_\eta \rangle $.

In terms of higher spin theory, other chiral
primary with generic $(p,l)$
should correspond to the bound state of $(p + l)$ scalars.
However, according to
\cite{Chang:2012kt} (see \cite{Gaberdiel:2013vva} for the ${\cal N}=4$ holography on AdS$_3$),
a multi-particle state in higher spin theory may be regarded as a single-string state.
As argued above, dual superstring theory may be on AdS$_3 \times $M$_7$,
and currently there are only three explicit candidates as
M$_7 = ($S$^3 \times $S$^3 \times $S$^1)/\mathbb{Z}_2$ \cite{Yamaguchi:1999gb},
and M$_7 = $SU(3)$/$U(1), SO(5)$/$SO(3) \cite{Argurio:2000tg}.
For the latter two cases, space-time chiral primaries have been already studied
in \cite{Argurio:2000xm}, and
they were found to be labeled by two integers $q = (p+l )/2$.
Therefore, we can say from \eqref{BPS} that the BPS spectrum matches with the one discussed  in this subsection.
Actually, we can easily see that there are many other chiral primaries in our coset.
Our conjecture is that they are dual to multi-string states, but currently there is no strong evidence for the conjecture.%
\footnote{We would like to report several supports for the conjecture in a separate publication. See footnote 4 in \cite{Argurio:2000xm} for a related issue.}

Now the model has three chiral primaries with $q=1$ for $(p,l) = (2,0),(1,1),(0,2)$.
We can construct three marginal operators with $h=1$ from them
by acting ${\cal N}=3$ supercurrents to construct the singlet of so$(3)$ R-currents.
The deformations by these operators are exactly marginal, so the model has three moduli parameters, see, e.g., section 4 of \cite{Argurio:2000xm}.
In order to relate to superstring theory, higher spin symmetry
should be broken by these marginal deformations.
For the ${\cal N}=4$ holography on AdS$_3$, there is a marginal deformation \cite{Gaberdiel:2013vva}, which was shown to break higher spin symmetry in section 5 of \cite{Gaberdiel:2014yla}.

In this section we have proposed two kinds of dualities. One is between the coset \eqref{coset} with $k=N+M$ and with some fermions decoupled and higher spin gravity with extended supersymmetry. The other is between the Grassmannian coset \eqref{KScoset} with $k=N+M$ and a superstring theory. The Kazama-Suzuki model is constructed by gauging a su$(M)$ factor and recovering the decoupled fermions from the other coset \eqref{coset}. Therefore, we have a new relation between superstring theory and higher spin gravity, but after assigning a kind of U$(M)$ invariant condition and adding boundary fermions to the higher spin gravity. It was pointed out in \cite{Candu:2013fta} that we can include the singlet condition by changing boundary conditions of higher spin gauge fields. Moreover, it was argued in \cite{Gaberdiel:2014yla} that we need to add extra fermions localized at the boundary to the higher spin gravity in order to have a linear large ${\cal N}=4$ superconformal symmetry as an asymptotic symmetry. The linear large ${\cal N}=4$ symmetry is that of CFT dual to superstring theory on AdS$_3 \times$S$^3 \times$S$^3 \times $S$^1$. Thus it looks natural to add boundary fermions to higher spin gravity so as to be dual to a superstring theory.

\section{Higher spin gravity at $\lambda=1/2$}
\label{gravity2}

In this section we will discuss the higher spin gravity when $\lambda=1/2$, how a natural truncation appears and extended supersymmetry at the linear level.

Let us first briefly remind ourselves of the $\mathcal{N}=2$ higher spin supergravity from  \cite{Prokushkin:1998bq}. The theory consists of scalar fields and fermions coupled to a gauge sector which is described by a $\text{shs}[\lambda] \oplus \text{shs}[\lambda]$ Chern-Simons theory. The Lie algebra $\text{shs}[\lambda]$ is generated by $y_\alpha$ $(\alpha =1,2)$ and $\hat k$ which anti-commutes with the $y_\alpha$. At $\lambda=1/2$ the fundamental commutator takes a particular simple form in
\begin{align}\label{eq:basiccommu}
     [y_\alpha , y_\beta]= 2 i \epsilon_{\alpha \beta}\big(1-(1-2\lambda)\hat k\big) = 2 i \epsilon_{\alpha \beta} \, .
\end{align}
The generators of $\text{shs}[\lambda]$ can now be written as symmetrized products of $(2s-2)$ $y_\alpha$, $V^{(s)\sigma}_m$
$(m < |s|)$ where the spin $s=1,3/2,2,5/2,\ldots$. Here $\sigma=\pm$ corresponds to whether we choose to include a $\hat k$ ($\sigma=+$) or not ($\sigma=-$). Further, we introduce Chan-Paton factors by considering generators of the form
\begin{align}\label{}
    V^{(s)\sigma}_m\otimes t_a
\end{align}
where $t_a$ are generators of gl$(M)$. Note that $V_0^{(1)+} \otimes \1_M=\1\otimes \1_M$ should be decoupled.

At $\lambda=1/2$ the operator $\hat k$ does not get generated by the basic commutator \eqref{eq:basiccommu}. As described in \cite{Prokushkin:1998bq} we thus have a special truncation for $\lambda=1/2$ via the automorphism taking $\hat k\mapsto-\hat k$, with corresponding involutive symmetry $\zeta$ taking $\zeta[W(\hat k)]=W(-\hat k)$, where $W$ is the general gauge field. Applying this to the case with GL$(M)$-extended symmetry (and doing no other truncations i.e. $\alpha=\beta=0$ in the terminology of \cite{Prokushkin:1998bq}), our generators are $V^{(s)+}\otimes t_a$, i.e. with no $\hat k$.

This reduces our original $\mathcal{N}=2$ supersymmetry generated by $G^\pm\propto V^{(3/2)+}\otimes \1_M\pm V^{(3/2)-}\otimes \1_M$ to the $\mathcal{N}=1$ symmetry generated by the sum of the two generators $G^++G^-$ i.e. by $y_\alpha \otimes t_a$. Finally, also the matter states should be reduced, and will only contain fields with no $\hat k$.

We thus have gl$(M)$ extended $\mathcal{N}=1$ higher spin supersymmetry where at first sight we can maximally have an $\mathcal{N}=1$ supersymmetry algebra.\footnote{For notation we call an algebra only consisting of spin one, spin 3/2 and the Virasoro generators for a supersymmetry algebra. A superalgebra can have higher spin fields, and other spin two content. Further, in this section we only consider what happens in the bulk, not the actual asymptotic symmetries which will be discussed in the next section.} Including more than one spin 3/2 generator will generically generate higher spin operators via (anti-)commutation relations.
It is, however, possible to find an extended supersymmetry algebra in certain cases. As stated in \cite{Prokushkin:1998bq} if we consider the case where $M=2^{p/2}$ then we have a so$(p)$-extended supersymmetry algebra, i.e. the generators form osp$(p|2)$. To see this, let $\phi^I$, $I=1,\ldots, p$ be the Clifford elements generating gl$(2^{p/2})$ with basic anti-commutator
\begin{align}\label{eq:Cliffordanticom}
  \{\phi^I,\phi^J\}=2\delta^{IJ} \ .
\end{align}
The so$(p)$ algebra is then generated by
\begin{align}\label{eq:Clifrota}
  M^{IJ}=[\phi^I,\phi^J]\ ,
\end{align}
and the global modes of the supercharges rotating under this group are
\begin{align}\label{eq:bulksupercharge}
  Q_\alpha^I =y_\alpha \otimes \phi^I\ ,
\end{align}
and the global modes of the stress-energy tensor is
\begin{align}\label{}
  T_{\alpha\beta}=\frac12\{y_\alpha,y_\beta\}\otimes 1_M\ .
\end{align}
Indeed we see that
\begin{align}\label{}
  \{Q^I_\alpha,Q^J_\beta\}=\delta^{IJ}T_{\alpha\beta}+i\varepsilon_{\alpha\beta}M^{IJ}
\end{align}
by using that
\begin{align}\label{}
  \{y_\alpha\otimes t_a,y_\beta\otimes t_b\}=T_{\alpha\beta}\{t_a,t_b\}+i\varepsilon_{\alpha\beta}[t_a,t_b]\ .
\end{align}

\subsection{D$(2,1|\alpha)$}

Let us briefly note the relation to the D$(2,1|\alpha)$ algebra discovered in \cite{Gaberdiel:2013vva} for the case $M=2$, but which can be embedded for all even $M$-values. At $\lambda=1/2$ we have $\alpha=\lambda/(1-\lambda)=1$ and D$(2,1|1)$ is simply osp$(4|2)$. The supercharges of this algebra however depend on the existence $\hat k$ and osp$(4|2)$ does not survive the projection. However, a sub-superalgebra osp$(2|2)$ (i.e. $\mathcal{N}=2$) does survive the projection, and it has the supercharges
\begin{align}\label{eq:twogenerators}
  y_\alpha\otimes\begin{pmatrix}
                   0 & 1 \\
                   0 & 0 \\
                 \end{pmatrix}\ ,\qquad   y_\alpha\otimes\begin{pmatrix}
                   0 & 0 \\
                   1 & 0 \\
                 \end{pmatrix}\
\end{align}
which are $i\hat k G^{++}$ and $i\hat k G^{--}$ in the notation of \cite{Gaberdiel:2013vva} up to normalization.\footnote{We have here multiplied the supercharges in \cite{Gaberdiel:2013vva} with $i\hat k$, which does not change the anti-commutators, to get $\hat k$ independent operators.} This is indeed the so(2)-extended algebra that we found in the last section.

\subsection{From so$(p)$ to so$(p+1)$-extended supersymmetry algebra}

There is, however, a surprise when we consider the case $M=2$, i.e. $p=2$. Inside the osp$(4|2)$ superalgebra we also have an osp$(3|2)$ algebra which is preserved by our reduction. To get this we simply need to add the generator
\begin{align}\label{}
  y_\alpha\otimes\begin{pmatrix}
                   1 & 0 \\
                   0 & -1 \\
                 \end{pmatrix}
\end{align}
to the two generators in \eqref{eq:twogenerators}. In the notation of \cite{Gaberdiel:2013vva} this is $i\hat k (G^{+-}+G^{-+})$.

The proposal is now that this is a general feature, i.e. the osp$(p|2)$ superalgebra of \cite{Prokushkin:1998bq} actually is a subalgebra of a osp$(p+1|2)$ superalgebra. To see this consider gl$(2^{p/2})$ in the tensor representation where the generators are
\begin{align}\label{}
 \sigma^{\delta_1}\otimes\cdots\otimes\sigma^{\delta_{p/2}}\ ,\qquad \delta_n=0,1,2,3 \ .
\end{align}
The $p$ Clifford elements generating gl$(2^{p/2})$ are
\begin{align}\label{}
  \phi^I\equiv\phi^{\delta a}= \underbrace{\sigma^3\otimes\cdots\otimes\sigma^3}_{a-1}\otimes\sigma^\delta\otimes \1_2 \otimes\cdots\otimes \1 _2    \ ,\qquad a=1,\ldots,p/2, \ \delta=1,2 \ .
\end{align}
Indeed these fulfill \eqref{eq:Cliffordanticom} and the so$(p)$ algebra is generated by
\begin{align}\label{}
  [\phi^{\delta a},\phi^{\delta' a'}]\in\textrm{span}\{&1\otimes\cdots\otimes \1_2 \otimes\sigma^3\otimes \1_2\otimes\cdots\otimes \1_2 \ , \\
  &1\otimes\cdots\otimes1\otimes\sigma^{\delta_1}\otimes\sigma^3\otimes\cdots\otimes\sigma^3\otimes\sigma^{\delta_2}\otimes \1\otimes\cdots\otimes \1_2 |\delta_1,\delta_2=1,2\}\ , \nonumber
\end{align}
for which we count $p/2+4\binom{p/2}{2}=p(p-1)/2$ elements. The idea is that we can add another generator
\begin{align}\label{}
  \phi^{p+1}=\sigma^3\otimes\cdots\otimes\sigma^3\ .
\end{align}
Even with this generator we still have the relation \eqref{eq:Cliffordanticom} which ensures that the only spin two generator generated by the supercharges $y_\alpha\otimes \phi^I$ with $I=1,\ldots,p+1$ is the Virasoro tensor. For the commutators of the $\phi^I$s we now have $p$ extra elements
\begin{align}\label{}
  [\phi^{p+1},\phi^{\delta' a'}]\in \underbar{p}=\textrm{span}\{\1_2 \otimes\cdots\otimes \1_2 \otimes\sigma^\delta\otimes\sigma^3\otimes\cdots\otimes\sigma^3|\delta=1,2\}\
\end{align}
 which is a subalgebra and forms a dimension $p$ irreducible representation of so$(p)$. This is similar to the decomposition $\textrm{so}(p+1)=adj(\textrm{so}(p))+\underbar{p}$ and we thus expect this to be so$(p+1)$. Finally, we need to check that the supercharges transform in the vector representation of so$(p+1)$. Here we see that $y_\alpha\otimes \phi^{p+1}$ is not rotated by so$(p)$, and under the commutator with  $\underbar{p}$ it is taken to $y_\alpha\otimes \phi^{\sigma a}$, and finally the commutator with $\underbar{p}$ takes $y_\alpha\otimes \phi^{\sigma a}$ into $y_\alpha\otimes \phi^{p+1}$. We thus have the global modes of a so$(p+1)$-extended supersymmetry algebra.

\section{Orbifold CFT}\label{sec:orbi}

In the earlier sections, we have suggested CFT duals of the higher spin supergravity at $\lambda=1/2$ in terms of cosets at critical levels where a free fermion description is possible. In this section, we will take different approach. We will start from the CFT dual to the untruncated ${\cal N}=2$ higher spin supergravity with $M \times M$ matrix valued fields and then perform an orbifold truncation of this CFT dual to the anti-automorphism truncation by $\zeta$ carried out on the bulk side in the last section.

We thus start from the modified coset Grassmannian CFT
\begin{align}\label{eq:dualcft}
 \frac{\text{su}(N+M)_k \oplus \text{so}(2NM)_1}{\text{su}(N)_{k+M} \oplus
\text{su}(M)_{k+N} \oplus \text{u}(1)_\kappa}\times \text{su}(M)_{k+N}
\end{align}
which we proposed in \cite{Creutzig:2013tja} to be dual to the ${\cal N}=2$ higher spin supergravity in the large $N$ limit with fixed 't Hooft parameter
\begin{align}\label{eq:thooft2}
  \lambda=\frac{N}{N+k+M}\ .
\end{align}
On the bulk side we generally have a map which takes $\hat k\mapsto -\hat k$ and $\lambda \mapsto 1-\lambda$ and at $\lambda=1/2$ this becomes the anti-automorphism $\zeta$. On the CFT side, considering only the Grassmannian coset, we generally have the level-rank duality exchanging $k$ and $N$ which likewise takes the 't Hooft parameter $\lambda \mapsto 1-\lambda$. However, on the bulk side we see that the $\mathcal{N}=2$ supercharges are being exchanged by $\hat k\mapsto -\hat k$, whereas they are preserved under the level-rank duality, see \cite{Kazama:1988qp}. We thus propose that on the CFT side the dual transformation is given by level-rank duality together with conjugation and we will also denote this by $\zeta$. When $k=N$ i.e. $\lambda=1/2$ this becomes a symmetry of the theory, and to compare with the bulk side we should consider the orbifold with respect to this $\mathbb{Z}_2$ automorphism. Note that $k=N$ is a non-critical level, but is special in the sense that the level-rank duality is an automorphism.

\subsection{Supersymmetry algebra reduction}

In this subsection we will examine how the suggested automorphism acts on the supersymmetry algebra.

Let us first consider the supercurrents. We use the vertex operator representation as in \cite{Kazama:1988qp} for $\text{su}(N+M)$. We denote the basis of the root vectors in the su$(N)$ direction by $e_i$ with $i=1,\ldots, N$, and in the su$(M)$ direction by $e_{N+A}$ with $A=1,\ldots, M$.%
\footnote{Within this section we use the notation in \cite{Creutzig:2013tja}, which is actually
different from the one used in the rest of paper. With the notation we can easily borrow the
results from the paper.}
The weight corresponding to the upper right diagonal in su$(N+M)$ then take the form
\begin{align}\label{}
  \bar\alpha^{i,A}=e_i-e_{N+A} \ .
\end{align}
We introduce $k(N+M)$ free scalars $\phi^I_K$ with $I=1,\ldots, N+M$ and $K=1,\ldots,k$ having OPEs
\begin{align}\label{}
  \phi^I_K(z)\phi^{I'}_{K'}(0)\sim-\delta^{II'}\delta_{KK'}\ln|z|^2\ .
\end{align}

We then have the following vertex operator representation of the currents in the off-diagonal blocks of $\text{su}(N+M)$ (in the notation from \cite{Creutzig:2013tja} where the generators of su$(N+M)$ are denoted $t_{IJ}$)
\begin{align}\label{}
  {J^A}_i=&J_{t_{i,N+A}}=\sum_K e^{i\bar\alpha^{i,A}\cdot\phi_K}=\sum_K e^{i\phi^i_K-i\phi^{N+A}_K}\ ,\\
  {J^i}_A=&J_{t_{N+A,i}}=\sum_K e^{-i\bar\alpha^{i,A}\cdot\phi_K}=\sum_K e^{-i\phi^i_K+i\phi^{N+A}_K}\ ,
\end{align}
where co-cycles are suppressed. Of course, the corresponding Sugawara tensor can only be the standard free scalar Virasoro tensor for $k=1$, but we do not want to make calculations, but just see the action of the level-rank duality like in \cite{Kazama:1988qp}.

We also need to bosonize the fermions, introducing $MN$ free scalars
\begin{align}\label{}
  {\psi^A}_i=\sqrt{k+N}e^{-i\phi^A_i}\ ,\qquad {{\bar\psi}^i}_A=\frac1{\sqrt{k+N}}e^{i\phi^A_i}\ .
\end{align}
Then the u$(M)$ extended supercharges are (see \cite{Creutzig:2013tja})
\begin{align}\label{eq:supercharges}
  G^+[t_{N+A,N+B}]=&\sqrt{2}{\bar\psi^i}_A {J^B}_i=\frac{\sqrt{2}}{\sqrt{k+N}}\sum_{i,K} e^{i\phi^A_i+i\phi^i_K-i\phi^{N+B}_K}\ ,\\
  G^-[t_{N+A,N+B}]=&\frac{\sqrt{2}}{k+N} {\psi^B}_i{J^i}_A=\frac{\sqrt{2}}{\sqrt{k+N}} \sum_{i,K} e^{-i\phi^B_i-i\phi^i_K+i\phi^{N+A}_K}\ .
\end{align}

The level-rank duality acts by exchanging the fields $\phi^A_i$ and $\phi^{N+A}_K$ with a sign change, and transposing $\phi^i_K$. We thus suggest the $\zeta$ takes the form (with extra signs compared to the level-rank duality)
\begin{align}\label{}
  \zeta(\phi^A_i)=\phi^{N+A}_{K=i}\ , \qquad \zeta(\phi^i_K)=-\phi^K_i \ .
\end{align}
We then see that
\begin{align}\label{}
  \zeta\big(G^+[t_{N+A,N+B}]\big)=G^-[t_{N+A,N+B}]\ ,
\end{align}
and in the orbifold we should only keep the sum
\begin{align}\label{}
  G^+[t_{N+A,N+B}]+G^-[t_{N+A,N+B}]\ ,
\end{align}
which on the bulk side exactly correspond to the spin-3/2 operators independent of $\hat{k}$.

Let us now consider the spin-one operators. Using that the su$(M)$ currents in the vertex operator representation has the form
\begin{align}\label{}
  J_{t_{N+A,N+B}}=\sum_K:e^{i\phi^{N+A}_K}e^{-i\phi^{N+B}_K}:\ ,
\end{align}
we see that the u$(M)$ extended u(1) current has the form (being careful with signs)
\begin{align}\label{}
  J^{(1)-}[t_{N+A,N+B}]=&-\lambda J_{t_{N+A,N+B}}+\lambda\frac{k}{N}:{\bar\psi^i}_A{\psi^B}_i:\nonumber\\
  =&-\lambda\sum_K:e^{i\phi^{N+A}_K}e^{-i\phi^{N+B}_K}+\lambda\frac{k}{N}:\sum_i:e^{i\phi^A_i}e^{-i\phi^B_i}:\nonumber\\
  \stackrel{k=N}{=}&\,-\frac12\sum_K:e^{i\phi^{N+A}_K}e^{-i\phi^{N+B}_K}+\frac12:\sum_i:e^{i\phi^A_i}e^{-i\phi^B_i}:\ .
\end{align}
We see that at $\lambda=1/2$ we have
\begin{align}\label{}
  \zeta\big(J^{(1)-}[t_{N+A,N+B}]\big)=-J^{(1)-}[t_{N+A,N+B}]\ ,
\end{align}
which fits perfectly with $J^{(1)-}[t_{N+A,N+B}]$ being dual to $\frac12(1-2\lambda+\hat k)\otimes t_{N+A,N+B}=\frac12\hat k\otimes t_{N+A,N+B}$ on the bulk side. The current dual to $\1\otimes t_a$ on the other hand takes the form
\begin{align}\label{}
  J^{(1)+}[t_{N+A,N+B}]=&J_{t_{N+A,N+B}}+:{\bar\psi^i}_A{\psi^B}_i:\nonumber\\
    =&\sum_K:e^{i\phi^{N+A}_K}e^{-i\phi^{N+B}_K}:+\sum_i:e^{i\phi^A_i}e^{-i\phi^B_i}:
\end{align}
and is invariant under $\zeta$ as expected.

\subsection{Check of OPEs}

We will now consider the OPEs of the supercurrents using the results of \cite{Creutzig:2013tja} (obtained via the methods of \cite{Creutzig:2012xb,Moradi:2012xd,Ammon:2011ua}) where these OPEs were successfully compared on both sides of the duality. The OPEs of the the supercurrents for general $\lambda$ were in the large $N$ limit calculated to be
\begin{align}\label{eq:supercurrentopebulk}
    G^{-}[t_a](z)G^{+}[t_b](0)\sim&\frac{2 c g_{ab}}{3M z^3}
    +\frac{2A_{ab}}{z^2}\\
    &+\frac{{s_{ab}}^cJ^{(2)+}[t_c]-{f_{ab}}^c J^{(2)-}[t_c]+\partial A_{ab}+\frac{3M}{c}{A^c}_bA_{ac}+\mathcal{O}(t_a,t_b)}{z}\nonumber
\end{align}
where
\begin{align}\label{}
    A_{ab}=\lambda(1-\lambda){f_{ab}}^cJ^{(1)+}[t_c]+((\tfrac12-\lambda){f_{ab}}^c-\tfrac12{s_{ab}}^c)J^{(1)-}[t_c]  \ ,
\end{align}
and $\mathcal{O}(t_a,t_b)$ are derivative terms that can appear when choosing a normal ordering for the non-linear term $\frac{3M}{c}{A^c}_bA_{ac}$. The current $J^{(2)+}[t_c]$ is dual to $V^{(2)+}_m\otimes t_c$, and $J^{(2)+}[\1_M]$ is the stress-energy tensor up to a terms $\frac{3M}{2c}g^{ab}J^{(1)-}[t_a]J^{(1)-}[t_b]$. The generators $t_a$ are those of su$(M)$ and ${f_{ab}}^c$ are the corresponding structure constants and ${s_{ab}}^ct_c=\{t_a,t_b\}$ and finally $g_{ab}=\tr{t_at_b}$.

Inserting the suggested supercharges $Q^I\equiv G^{+}[\phi^I]+G^{-}[\phi^I]$ we find using \eqref{eq:Cliffordanticom} and \eqref{eq:Clifrota} (ignoring the terms in $\mathcal{O}(t_a,t_b)$)
\begin{align}
    Q^I(z)Q^J(0)\sim&\frac{4 c\delta^{IJ}}{3 z^3}
    +\frac{2A_{\phi^I\phi^J}-2A_{\phi^J\phi^I}}{z^2}\\
    &+\frac{4\delta^{IJ}J^{(2)+}[\1_M]+\partial A_{\phi^I\phi^J}-\partial A_{\phi^J\phi^I} +\frac{3M}{c}{A^c}_{\phi^J}A_{\phi^I c}+\frac{3M}{c}{A^c}_{\phi^I}A_{\phi^J c}}{z}\nonumber
\end{align}
which gives
\begin{align}
    Q^I(z)Q^J(0)\sim&\frac{8 k_{CS}\delta^{IJ}}{ z^3}
    +\frac{J^{(1)+}[M^{IJ}]}{z^2}+\frac{4\delta^{IJ}J^{(2)+}[\1_M]+\frac12\partial J^{(1)+}[M^{IJ}] }{z}\\
    &+\frac{\frac{M}{4k_{CS}}\big(\frac14{{f^c}_{\phi^J}}^d {f_{\phi^I c}}^eJ^{(1)+}[t_d]J^{(1)+}[t_e]+{{s^c}_{\phi^J}}^d {s_{\phi^I c}}^eJ^{(1)-}[t_d]J^{(1)-}[t_e]\big)}{z}\nonumber \ .
\end{align}

For finite $N,k$ the non-linear terms produced here would firstly give $J^{(1)-}J^{(1)-}$ terms which are unwanted. Secondly, when the non-linear terms act on the supercurrents they would generate spin $3/2$ currents which are not of the $Q^I$ form, and these will in turn generate currents of all spins. We thus do not get a supersymmetry algebra. However, if we in the large $N$ limit only keep operators that scales such that their central term goes like $N$ (as in \cite{Creutzig:2013tja}), we can safely ignore the non-linear terms. In conclusion, in the large $N$ limit for $M=2^{p/2}$ we have an so$(p+1)$ extended supersymmetry algebra, and this is invariant under $\zeta$. Actually, we could also see this as a large $N$ limit of Knizhnik's so$(N+1)$ extended superconformal algebra \cite{Knizhnik:1986wc}.

\subsection{State reduction}

Finally, we will consider the coset states dual to the bulk matter fields. The level-rank duality of the coset states was investigated in \cite{Naculich:1997ic}. Let us denote the states of the coset
\begin{align}
 \frac{\text{su}(N+M)_k\oplus \text{so}(2NM)_1 }{\text{su}(N)_{k+M} \oplus \text{su}(M)_{k+N}
 \oplus \text{u}(1)_{NM(N+M)(N+M+k)}} \ ,
\end{align}
by (similar to section \ref{sec:superstrings} above)
\begin{align}\label{}
    (\Lambda_{N+M},\omega;\Lambda_N,\Lambda_M,m)\ .
\end{align}
The conformal dimension is found by summing the nominator factors and subtracting the parts from the denominator:
\begin{align}
 h = n + h^{M+N,k}_{\Lambda_{M+N}} + h_\omega^{2MN} - h^{N,k+M}_{\Lambda_N} - h^{M,k+N}_{\Lambda_M}
 - h_m \ ,
\end{align}
for some integer level $n$. We will only consider the NS sector where $\omega=0,2$ and the selection rules then take the form
\begin{align} \label{selection}
&m =  N |\Lambda_{N+M}| - (N+M) |\Lambda_N| - N (N+M) a \,  , \\
&m =  -M |\Lambda_{N+M}| + (N+M) |\Lambda_M| - M (N+M) b \nonumber \ .
\end{align}
This defines $a$ and $b$. Note here that $m$ is here defined modulo $NM(N+M)(N+M+k)$.

We also remember that we have two types of field identifications
\begin{align}\label{}
  J_1[(\Lambda_{N+M},\omega;\Lambda_N,\Lambda_M,m)]=&(\sigma(\Lambda_{N+M}),\sigma^M(\omega);\sigma(\Lambda_N),\Lambda_M,m-M(N+M+k))\ ,\nonumber\\
  J_2[(\Lambda_{N+M},\omega;\Lambda_N,\Lambda_M,m)]=&(\sigma(\Lambda_{N+M}),\sigma^N(\omega);\Lambda_N,\sigma(\Lambda_M),m+N(N+M+k))\ ,\label{eq:fieldid}
\end{align}
where $\sigma$ cycles the Dynkin indices, $a_i(\sigma(\Lambda))=a_{i-1}(\Lambda)$, and exchanges $\omega=0,2$.

The level-rank duality, exchanging $k$ and $N$ in the coset, then acts as follows \cite{Naculich:1997ic}
\begin{align}\label{}
  (\Lambda_{N+M},\omega;\Lambda_N,\Lambda_M,m)\mapsto (\Lambda^t_{N},\sigma^u(\omega);\Lambda^t_{N+M},\sigma^v(\bar{\Lambda}_M),\tilde m) \ ,
\end{align}
where we need to define $u,v$ and $\tilde m$. $v$ is determined as
\begin{align}\label{}
  v\equiv-a \mod (k,M)\ .
\end{align}
We can now determine an integer $s=0,\ldots,M/(k,M)-1$ (and the integer $t$) uniquely by the equation
\begin{align}\label{}
  -ks/(k,M)+Mt/(k,M)=(a+v)/(k,M) \ .
\end{align}
Then $u$ is determined modulo 2 by
\begin{align}\label{}
  u=|\Lambda_{N+M}|-|\Lambda_{N}|+(k-N)a+k(k+M)s=|\Lambda_{N+M}|-|\Lambda_{N}|+k(k+M)s\ ,
\end{align}
where we in the last equality used $k=N$. Further, let $\tilde a$ be the ``$a$'' from the selection rules of the level-rank dualized weight. Then $\tilde a$ is determined by $s$ as
\begin{align}\label{}
  \tilde a=a+(N+M+k)s \ .
\end{align}
Finally, $\tilde m$ is the determined from its selection rule, or alternatively compared to the selection rule for $m$ by
\begin{align}\label{}
  \frac{m-\tilde m}{N+M+k}\equiv u \mod (2)M \,
\end{align}
where we have modulo $M$ for $M$ even, and modulo $2M$ for $M$ odd.

At last, our $\zeta$ should also contain conjugation
\begin{align}\label{}
    \zeta[(\Lambda_{N+M},\omega;\Lambda_N,\Lambda_M,m)]= (\overline{\Lambda^t_{N}},\sigma^u(\omega);\overline{\Lambda^t_{N+M}},\overline{\sigma^v(\bar{\Lambda}_M)},-\tilde m)\ .
\end{align}

We can now use this on the fundamental states found in \cite{Creutzig:2013tja}. Using \eqref{eq:supercharges} (without going to the vertex operator representation) we see that these states have the supersymmetry structure displayed in table \ref{tab:states}.
\begin{table}\begin{center}
\begin{tabular}{|c|c|c|c|c|}
  \hline
  Chiral state & h &  & superpartner & h \\\hline
  $(\bar f,0;0,\bar f,-N)$ & $\frac{\lambda}{2}=\frac{1}{4}$ & $\stackrel{G^-}{\mapsto}$ & $(\bar f,2;0,\bar f,-N)$  & $\frac{1+\lambda}{2}=\frac{3}{4}$ \\\hline
  $(0,2;\bar f,f,N+M)$ & $\frac{1-\lambda}{2}=\frac{1}{4}$ & $\stackrel{G^-}{\mapsto}$  & $(0,0;\bar f,f,N+M)$ & $\frac{2-\lambda}{2}=\frac{3}{4}$ \\
  \hline
  Anti-chiral state & h &  & superpartner & h \\\hline
    $(f,0;0, f,N)$ & $\frac{\lambda}{2}=\frac{1}{4}$ & $\stackrel{G^+}{\mapsto}$ & $(f,2;0,f,N)$  & $\frac{1+\lambda}{2}=\frac{3}{4}$ \\\hline
  $(0,2; f,\bar f,-N-M)$ & $\frac{1-\lambda}{2}=\frac{1}{4}$ & $\stackrel{G^+}{\mapsto}$  & $(0,0;f,\bar f,-N-M)$ & $\frac{2-\lambda}{2}=\frac{3}{4}$ \\
  \hline
\end{tabular}
\end{center}
\caption{The fundamental states in the large $N$ limit and their
relations under the supersymmetry algebra.}
\label{tab:states}
\end{table}
Using $\zeta$ on these states, we see that it connects the supermultiplets by taking $\lambda\mapsto 1-\lambda$:
\begin{align}\label{}
    \zeta[(\bar f,0;0,\bar f,-N)]=& (0,2; f,\bar f,-N-M)\ ,\nonumber \\
    \zeta[(0,2;\bar f,f,N+M)]=& (f,0;0, f,N)\ .
\end{align}
In order to get the first equality, one has to use the field identifications \eqref{eq:fieldid}. We note that the weight for the su$(M)$ factor is kept invariant, which is good since this is the factor we want to remove from the denominator, see \eqref{eq:dualcft}.

To compare with the bulk side, we should apply $\zeta$ on the left- and right-moving side at the same time. Thus, for the fundamental fields, we should only keep linear combinations such as
\begin{align}\label{}
    (\bar f,0;0,\bar f,-N)\otimes (f,0;0,f,N)+(0,2; f,\bar f,-N-M)\otimes (0,2; \bar f, f,N+M)\ ,
\end{align}
and this directly corresponds to the $\hat k$ independent state on the bulk side.

\subsection{Partition function}

A naive analysis on both sides of the duality gives a match of the partition function. On the bulk side, we should constrain to states independent of $\hat k$ -- both for the higher spin fields and for the matter fields. This means that the partition function is the square root of the one obtained in \cite{Creutzig:2013tja}, except for the spin-one factor, see section \ref{gravity1}. On the CFT side the same thing should be happening when we assume that $\zeta$ does not change the fusion rules and we leave out twisted section. This is so, since according to the analysis of \cite{Gaberdiel:2010pz} the fusion of the different multiplets will reveal null vectors in such a way that we just get products of infinite fusions of the fundamental states. When we remove some of these states, we should simply remove the corresponding products from the partition function.

Note that the method used in this section was very different from the earlier section and suggests that in the large $N$ limit we have a relation like
\begin{align}\label{}
         \frac{\text{su}(N+M)_N\oplus \text{so}(2NM)_1 }{\text{su}(N)_{N+M} \oplus \text{su}(M)_{2N}
 \oplus \text{u}(1)_{NM(N+M)(2N+M)}}\times\text{su}(M)_{2N} \bigg /\zeta\equiv\nonumber \\
  \frac{\text{su}(N'+M/2)_{N'+M/2}\oplus \text{so}(N'M)_1 }{\text{su}(N')_{N'+3M/2} \oplus \text{su}(M/2)_{2N'+M/2}
 \oplus \text{u}(1)_{\kappa}}\times\text{su}(M/2)_{2N'+M/2}\ .
\end{align}
Where the central charges will match in the large $N$ limit if $N'=2N$. For finite $N$ not even the central charge will match, and corrections to this relation will certainly be necessary.

\section{Conclusion and outlook}
\label{conclusion}

In this paper, we have proposed a duality between higher spin gravity with ${\cal N}=p+1$ extended
supersymmetry in \cite{Prokushkin:1998bq} and the bosonic coset model \eqref{bcoset} with $k=N$ and
with a non-diagonal modular invariant.
In order to obtain support for the duality, we have compared spectrum and symmetry.
The partition function of the coset model \eqref{bcoset} with $k=N$ has been
computed in \eqref{eq:CFTpt} in
the large $N$ limit, and it reproduces the gravity partition function in the form of \eqref{eq:partrewritten}.
The symmetry generators of the coset model at low spins have been constructed explicitly by
making use of a (conjectured) relation \eqref{eq:compboscoset}.
They are compared with low spin gauge fields in the bulk theory.
We can also relate the bosonic model \eqref{bcoset} with
the supersymmetric form of the coset in \eqref{coset} with $k=N \pm M$ and with fermions decoupled
as in \eqref{eq:compsupercoset}. In particular, we have computed the partition function
of the coset \eqref{coset} with $k=N-M$ at the large $N$ limit, and shown the agreement with
the gravity partition function \eqref{eq:partrewritten} up to contributions from free fermions.

We also considered the truncation done in the higher spin gravity in detail and performed a similar truncation on the CFT side to yield another dual theory in the large $N$ limit.
Since this is a truncation by an automorphism, it would be interesting to investigate the dual CFT as an orbifold model and to see how to obtain twisted sectors in the duality. Further,
also other truncations are possible on the gravity side at $\lambda=1/2$. Some of these preserve the original $\mathcal{N}=2$ duality, but reduce the matrix algebra of the Chan-Paton factors
to o$(N)$ or usp$(N)$. One can speculate that the duals are related to the remaining cosets in the Kazama-Suzuki table based on these bosonic algebras, i.e.
\begin{align}\label{}
  \mathcal{N}=2\textrm{ supercoset based on }\frac{\textrm{so}(2N)}{\textrm{su}(N)\oplus \textrm{u}(1)}
\end{align}
and
\begin{align}\label{}
  \mathcal{N}=2\textrm{ supercoset based on }\frac{\textrm{sp}(N)}{\textrm{su}(N)\oplus \textrm{u}(1)}\ .
\end{align}
We will leave this for future work.

A special motivation for studying this duality is to find a relation between 3d higher spin theory
and superstring theory via 2d CFT just like the ABJ triality in \cite{Chang:2012kt}.
In order to see relations to superstring theory, a su$(M)$ factor in the
supersymmetric form \eqref{coset} is gauged, and this leads to the Grassmannian Kazama-Suzuki coset \eqref{KScoset}.
We have shown that the coset with $k=N+M$ has ${\cal N}=3$ superconformal symmetry.
With the large supersymmetry, only three explicit candidates are known for the target space of
dual superstring theory, such as AdS$_3 \times $M$_7$ with M$_7 = ($S$^3 \times $S$^3 \times$S$^1 )/\mathbb{Z}_2$, SU$(3)/$U$(1)$ and SO$(5)/$SO$(3)$.
We have checked the relation to the Kazama-Suzuki model  by comparing
BPS states and marginal deformations for the latter two cases.
For the first case, we would like to study its spacetime chiral primaries and moduli parameters as a
future work. Detailed investigations on chiral primary states of critical level models would be
also important, see \cite{Gopakumar:2012gd,Isachenkov:2014zua}.

It is necessary to make the relation between superstring theory and
the Kazama-Suzuki model \eqref{KScoset} with $k=N+M$ more concrete.
We have shown that there are three moduli parameters in the critical level model, and
we would like to study the dependence of the deformation parameters.
In particular, we have to show that the higher spin symmetry is broken by these marginal deformations, see \cite{Gaberdiel:2013jpa,Gaberdiel:2014yla}.
The ${\cal N}=3$ superconformal field theories
with superstring duals
have not been specified yet.
Even so, there are several works on the closely related case, namely, the large ${\cal N}=4$ holography
\cite{Elitzur:1998mm,deBoer:1999rh,Gukov:2004ym}. For instance, it is argued recently in \cite{Tong:2014yna} that corresponding brane configuration would be useful to identify the dual CFT.
There might be several ${\cal N}=3$ superconformal field theories
with a superstring dual.
We expect that one of them is connected by a marginal deformation to the coset model \eqref{KScoset} at the critical level.
We would of course like to specify which ${\cal N}=3$ theory is connected to the coset model,
however this seems to require a more elaborate examination of non-BPS states.

\subsection*{Acknowledgements}

We are grateful to T.~Eguchi, K.~Ito, V.~Schomerus, Y.~Sugawara and S.~Yamaguchi for useful discussions.
The work of YH was supported in part by JSPS KAKENHI Grant Number 24740170. The work of PBR is funded by AFR grant 3971664 from Fonds National de la Recherche, Luxembourg, and partial support by the Internal Research Project GEOMQ11 (Martin Schlichenmaier), University
of Luxembourg, is also acknowledged. The work of TC is supported by NSERC grant number RES0019997.

\appendix

\section{Partition functions in the 't Hooft limit}
\label{PF}

In this appendix we give detailed computations on the partition functions of the coset models
in the 't Hooft limit.

\subsection{The bosonic coset}

As preparations we examine the properties of (super) Schur functions.
Let us define Schur functions $s_\alpha (x)$ as
\begin{align}
 s_\alpha (x) = \text{ch}_\alpha (X) = \sum_{T \in \text{Tab}_\alpha} \prod_{j \in T} X_{jj}
\end{align}
with $\text{Tab}_\alpha$ as the Young tableau of  shape $\alpha$.
Here we set $X_{jj} = x_{j+1}$ and $x=(x_1,x_2 , \ldots )$.
According to \cite{MacDonald}, we have a formula
\begin{align}
 \prod_{i,j=1}^\infty (1 - x_i y_j)^{-1} = \sum_{\alpha} s_\alpha (x) s_\alpha (y) \, ,
\end{align}
where the sum runs over all possible Young diagrams $\alpha$.
Introducing $M$ sets $x^{(A)},y^{(A)}$, we have (see appendix C of \cite{Creutzig:2013tja})
\begin{align}
\prod_{A,B=1}^M \prod_{i,j=1}^\infty (1 - x^{(A)}_i y^{(B)}_j) ^{-1} =
\sum_{\alpha} s_\alpha (x^{(1)},\ldots , x^{(M)}) s_\alpha (y^{(1)},\ldots , y^{(M)}) \, .
\end{align}
Using
\begin{align}
 s_\alpha (x ,y) = \sum_{\beta , \gamma} c^\alpha_{\beta \gamma } s_{\beta}(x)  s_{\gamma} (y)
\end{align}
repeatedly, we find
\begin{align}
 s_\alpha (x^{(1)},\ldots , x^{(M)})
 = \sum_{\alpha_1 , \ldots , \alpha_M} c^{\alpha}_{\alpha_1 \ldots  \alpha_M}
 \prod_{A=1}^M s_{\alpha_A} (x^{(A)}) \, ,
\end{align}
where we have used \eqref{cM}.
Setting
\begin{align}
 x^{(A)}_{i+1}  = q^{h+i} \, , \quad
 y^{(A)}_{i+1} = \bar q^{h+i}
\end{align}
for all $A=1, \ldots , M$, the one-loop partition function for a bulk real scalar field
can be written as
\begin{align}
 Z_\text{scalar} (q) =
\left( \prod_{i,j=0}^\infty \frac{1}{1 - q^{h+i} \bar q^{h+j}} \right)^{M^2}
 = \sum_\alpha  B^{h,M}_\alpha (q)  B^{h,M}_\alpha (\bar q)\, .
\end{align}
Here $B^{h,M}_\alpha (q) $ was defined in \eqref{BhM}.

Defining skew Schur functions  $s_{\alpha / \beta} (x) = \sum_{\gamma} c^{\alpha}_{\beta \gamma } s_{\gamma} (x) $, we have a formula \cite{MacDonald}
\begin{align}
 \sum_{\rho} s_{\rho / \gamma^t} (x) s_{\rho ^t/\mu} (y)
= \prod_{i,j} (1 + x_i y_j) \sum_{\tau} s_{\mu^t / \tau} (x) s_{\lambda/\tau^t} (y) \, .
\end{align}
Setting
\begin{align}
x=(x^{(1)} , \ldots , x^{(M)}) \, , \quad y=(y^{(1)}, \ldots , y^{(M)}) \, , \quad
x^{(A)}_{i+1} = y^{(A)}_{i+1} = q^{3/4 + i} \, ,
\end{align}
we find
\begin{align}
 \sum_\rho B^{3/4,M}_{\rho/\lambda^t} (q) B^{3/4,M}_{\rho^t /\mu} (q) =
  \left( \prod_{s=2}^\infty z_F^{(s) }\right)^{M^2}
  \sum_\beta B^{3/4,M}_{\mu^t/\beta^t} (q) B^{3/4,M}_{\lambda /\beta} (q) \, .
	\label{a6e}
\end{align}
Here we have defined $B_{\alpha/\beta}^{h,M} = \sum_{\gamma} c^\alpha_{\beta \gamma} B_\gamma^{h,M}$.
The above formula plays an important role to arrive at \eqref{chiM0} as we will see below.

Next let us define super Schur functions
\begin{align}
 s_\alpha (x|\xi) = \text{sch}_\alpha ({\cal X})
 = \sum_{T \in \text{STab}_\alpha} \prod_{j \in T} {\cal X}_{jj} (-1)^j \, ,
\label{sschur}
\end{align}
as in (A.6) of \cite{Candu:2012jq},
where $\text{STab}_\alpha$ denote the Young supertableau of shape $\alpha$.
Here
\begin{align}
 {\cal X}_{2i , 2i} = x_{i+1} \, , \quad {\cal X}_{2i +1 , 2i+1} = \xi_{i+1}
\end{align}
and $\xi = (\xi_1 , \xi_2 , \ldots)$.
The super Schur functions satisfy (see (A.8) of \cite{Candu:2012jq})
\begin{align}
 \prod_{i,j} \frac{(1-x_i \eta_j)(1-y_i \xi_j)}{(1-x_iy_j)(1-\xi_i \eta_j)}
 = \sum_{\alpha} s_\alpha (x|\xi) s_{\alpha} (y|\eta) \, .
 \label{scauchy}
\end{align}
Using
\begin{align}
 s_\alpha (x ,y |\xi , \eta) =
\sum_{\beta , \gamma}c^\alpha_{\beta \gamma}  s_\beta (x|\xi) s_{\gamma} (y|\eta) \, ,
\label{ssum}
\end{align}
repeatedly as above, we have
\begin{align}
 s_\alpha (x^{(1)},\ldots , x^{(M)} | \xi^{(1)}, \ldots , \xi^{(M)})
 = \sum_{\alpha_1 , \ldots , \alpha_M} c^{\alpha}_{\alpha_1 \ldots  \alpha_M}
 \prod_{A=1}^M s_{\alpha_A} (x^{(A)} | \xi^{(A)}) \, .
\end{align}
Setting
\begin{align}
 x^{(A)}_{i+1}  = q^{h+i} \, , \quad \xi^{(A)}_{i+1} = - q^{h + 1/2 + i} \, , \quad
 y^{(A)}_{i+1} = \bar q^{h+i}  , \quad \eta^{(A)}_{i+1} = - \bar q^{h + 1/2 + i}
\end{align}
for all $A=1, \ldots , M$, we find
\begin{align}
\left( \prod_{i,j=0}^\infty \frac{(1 + q^{h+1/2 +i} \bar q^{h + j}) (1 + q^{h + i} \bar q^{h+ 1/2 + j})}{(1 - q^{h+i} \bar q^{h+j})(1 - q^{h+1/2 + i} \bar q^{h+1/2 + j})} \right)^{M^2}
 = \sum_\alpha  {\cal B}^{h,M}_\alpha (q)  {\cal B}^{h,M}_\alpha (\bar q)\, ,
\end{align}
where ${\cal B}^{h,M}_\alpha (q)$ defined in \eqref{sBhM}.

With the help of (A.7) of \cite{Candu:2012jq}
\begin{align}
 s_\alpha (x | \xi) = s_{\alpha^t} (- \xi | - x) \, ,
\end{align}
we have
\begin{align}
 s_\alpha (x) = s_\alpha (x | 0) \, , \quad
s_{\alpha^t} (\xi) = s_\alpha (0 | - \xi) \, .
\end{align}
With \eqref{ssum}, this leads to
\begin{align}
 s_{\alpha} (x| - \xi) =
 \sum_{\beta , \gamma} c^\alpha_{\beta \gamma} s_{\beta} (x) s_{\gamma ^t} (\xi) \, .
\end{align}
Setting
\begin{align}
 x = (x^{(1)}, \ldots , x^{(M)}) \, , \quad \xi = (\xi^{(1)} , \ldots , \xi^{(M)} ) \, , \qquad
 x^{(A)}_{i+1} = q^{h+i} , \quad \xi^{(A)}_{i+1} = q^{h+1/2+i} \, ,
\end{align}
we find
\begin{align}
 {\cal B}^{h,M}_\alpha (q) = \sum_{\beta , \gamma} c^\alpha_{\beta \gamma}
 B^{h,M}_\beta(q) B^{h+1/2,M}_{\gamma^t} (q) \, .
\label{calBBB}
\end{align}
Using \eqref{calBBB}, we can show
\begin{align}
 \sum_{\gamma} B^{5/4,M}_{\alpha/\gamma} (q) B^{3/4,M}_{\gamma^t/\beta} (q)
  = \sum_{\gamma, \sigma ,\delta} c^{\alpha}_{\gamma \sigma}  c^{\gamma^t }_{\beta \delta} B^{5/4,M}_{\sigma} (q)
    B^{3/4,M}_{\delta} (q)
= \sum_{\gamma} c^{\alpha}_{\gamma \beta^t} {\cal B}^{3/4,M}_{\gamma^t} (q)
\label{a7e}
\end{align}
as in (A.7) of \cite{Beccaria:2013wqa}.

With the above preparations we can now derive \eqref{chiLambdaM} from \eqref{chiLambdaM0}
by closely following the analysis in appendix A of \cite{Beccaria:2013wqa}.
We can show that
\begin{align}
& \sum_{\Xi \in \Omega} q^{- \frac{|\Xi^l| + |\Xi^r| }{4}}
\sum_{\Psi} C^{(N) \Psi^*}_{\Phi \Xi^*} B^{1,M}_{\Psi^l} (q)  B_{\Psi^r}^{1,M} (q)
\\
&= \sum_{\Xi^l , \pi , \gamma} q^{- \frac{|\Phi^l| + |\Phi^r|}{4}}
 B_{\Phi^l/\gamma}^{5/4,M} (q) B_{(\Xi^l)^t/\pi}^{3/4,M} (q)
 B_{\Xi^l/\gamma}^{3/4,M} (q) B_{\Phi^r/\pi}^{5/4,M} (q) \, . \nonumber
\end{align}
Here we have used
$
 |\alpha | = \sum_{A = 1}^M |\alpha_A|
$ when $c^\alpha_{\alpha_1 \ldots \alpha_M}$ is non-zero.
With \eqref{a6e} we find
\begin{align}
 \chi_\Lambda^M (q) = \chi_0^M (q)
 \sum_{\Phi^l,\Phi^r , \gamma ,\delta}  q^{\frac{|\Lambda^l| + |\Lambda^r| - |\Phi^l| - |\Phi^r|}{4}}
 R^{(N)}_{\Lambda \Phi} C^{(N) (\gamma^t , \delta)}_{((\Phi^l)^t,0) (0,\Phi^r)}
 \mathcal{B}_{\gamma^t}^{3/4,M} (q) \mathcal{B}_{\delta^t}^{3/4,M} (q) \, ,
\end{align}
where we have utilized \eqref{a7e} and
$\chi^M_0 (q)$ is defined in \eqref{chiM0}.
As in \cite{Gaberdiel:2011zw,Beccaria:2013wqa} we assume
\begin{align}
 R^{(N)}_{\Lambda \Phi} \to
r_{\Lambda^l \Phi^l} r_{\Lambda^r \Phi^r} \, , \quad
 C^{(N) (\gamma^t , \delta)}_{((\Phi^l)^t,0) (0,\Phi^r)}
\to \delta^{\gamma^t}_{(\Phi^l)^t} \delta^{\delta}_{\Phi^r}
\label{assumption}
\end{align}
at the large $N$ limit, where $r_{\alpha \beta}$ are the restriction coefficients for gl$(\infty)_+$.
Using (C.33) of \cite{Creutzig:2013tja} (see (3.59) of \cite{Candu:2012jq} for $M=1$)
\begin{align}
 \mathcal{B}_\alpha^{0,M} (q) =
  \sum_{\beta} r_{\alpha \beta} \mathcal{B}_{\beta^t}^{1/2, M} (q) \, ,
\end{align}
we arrive at \eqref{chiLambdaM}.
For $M=1$ the expression reproduces (3.20) of \cite{Beccaria:2013wqa} except that the
Young diagrams are not transposed here. This is because we are
using the dual expression of the coset as in \eqref{bcoset}.

\subsection{The supersymmetric coset}

Let us move to the one-loop partition function of the supersymmetric coset \eqref{coset}.
For the case we need more properties of super Schur functions defined in \eqref{sschur}.
With \eqref{scauchy} and \eqref{ssum},
we obtain
\begin{align}
 \sum_{\rho}s_{\rho / \lambda} (x|\xi) s_{\rho / \mu} (y| \eta)
 = \prod_{i,j} \frac{(1-x_i \eta_j)(1-y_i \xi_j)}{(1-x_iy_j)(1-\xi_i \eta_j)}\sum_{\tau} s_{\mu/\tau} (x|\xi) s_{\lambda/\tau} (y|\eta)
\end{align}
by repeating the analysis in \cite{MacDonald}.
Using
\begin{align}
 s_{\mu/\tau}(x|\xi) = \sum_{\lambda} c^{\mu}_{\lambda \tau} s_{\lambda} (x|\xi)
 = \sum_\lambda  c^{\mu^t}_{\lambda^t \tau^t} s_{\lambda^t} (- \xi| - x)
= s_{\mu^t/\tau^t} (- \xi| - x) \, ,
\end{align}
we find
\begin{align}
  \sum_{\rho}s_{\rho / \lambda} (x|\xi) s_{\rho^t / \mu^t} (y| \eta)
 = \prod_{i,j} \frac{(1+x_i y_j)(1+ \xi_i \eta_j )}{(1+x_i \eta_j)(1+\xi_i y_j)}\sum_{\tau} s_{\mu/\tau} (x|\xi) s_{\lambda^t/\tau^t} (y|\eta) \, .
\end{align}
Setting
\begin{align}
 &x = (x^{(1)}, \ldots , x^{(M)}) \, , \quad
y =  (y^{(1)}, \ldots , y^{(M)}) \, , \quad
\xi = (\xi^{(1)}, \ldots , \xi^{(M)}) \, , \quad
\eta = (\eta^{(1)}, \ldots , \eta^{(M)}) \, \nonumber \\
&x^{(A)}_{i+1} =  y^{(A)}_{i+1} = q^{1/4+i} \, , \quad
\xi^{(A)}_{i+1} = \eta^{(A)}_{i+1} = - q^{3/4+i} \, ,
\end{align}
we have
\begin{align}
  \sum_{\rho}{\cal B}^{1/4}_{\rho / \lambda} (q) {\cal B}^{1/4}_{\rho^t / \mu^t} (q)
	= Y^M(q) \sum_{\tau} {\cal B}^{1/4}_{\mu/\tau} (q) {\cal B}^{1/4}_{\lambda^t/\tau^t} (q)
	\label{calBYM}
\end{align}
with
 $\mathcal{B}_{\alpha/\beta}^{h,M} = \sum_{\gamma} c^\alpha_{\beta \gamma} \mathcal{B}_\gamma^{h,M}$
and
\begin{align}
Y ^M (q) = \left( \prod_{s=2}^\infty z_B^{(s)} (q) z_F^{(s)} (q) \right)^{2M^2} ( z_B^{(1)} (q) )^{2M^2}
( z_F^{(1)} (q) )^{M^2} \, .
\end{align}

With the above properties we derive \eqref{schiM} from \eqref{schiM0}.
We first focus on the $M=1$ case and then move to generic $M \neq 1$ case.
As in the bosonic case, we obtain
\begin{align}
& \sum_{\Xi \in \Omega} q^{- \frac{|\Xi^l|+|\Xi^r|}{4}}
\sum_{\Psi} C^{(N)\Psi^*}_{\Phi \Xi^*} \text{sch}_{(\Psi^l)^t} ({\cal U}_{1/2})
 \text{sch}_{(\Psi^r)^t} ({\cal U}_{1/2}) \\
 & = \sum_{\Xi^l , \pi , \gamma}  q^{- \frac{|\Phi^l|+|\Phi^r|}{4}}
    \text{sch}_{(\Phi^l)^t/\gamma^t} ({\cal U}_{3/4}) \text{sch}_{\Xi^l/\pi^t} ({\cal U}_{1/4})
		\text{sch}_{(\Xi^l)^t/\gamma^t} ({\cal U}_{1/4}) \text{sch}_{(\Phi^r)^t/\pi^t} ({\cal U}_{3/4})
		\, .\nonumber
\end{align}
Here we have used $\text{sch}_{\alpha / \beta} (\mathcal{U}_h)= \sum_\gamma c^\alpha_{\beta \gamma} \text{sch}_\gamma (\mathcal{U}_h)$.
Defining
\begin{align}
\hat {\cal B}^{1/4,2}_{\gamma^t} (q)
	 = \sum_{\epsilon , \delta} c^{\gamma^t}_{\epsilon \delta^t}
	    \text{sch}_{\epsilon} ({\cal U}_{3/4})
			\text{sch}_{\delta} ({\cal U}_{1/4}) \, ,
\end{align}
we have
\begin{align}
 \sum_{\gamma} \text{sch}_{(\Phi^l)^t/\gamma^t} ({\cal U}_{3/4})
\text{sch}_{\gamma/\tau} ({\cal U}_{1/4})
 = \sum_{\gamma} c^{(\Phi^l)^t}_{\gamma^t \tau^t} \hat {\cal B}^{1/4,2}_{\gamma^t} (q) \, .
\end{align}
With \eqref{calBYM} we then arrive at
\begin{align}
 s \chi_\Lambda^1 (q) = \chi_0^2 (q) z_F^{(1)} (q)
 \sum_{\Phi^l,\Phi^r , \gamma ,\delta}  q^{\frac{|\Lambda^l| + |\Lambda^r| - |\Phi^l| - |\Phi^r|}{4}}
 R^{(N)}_{\Lambda \Phi} C^{(N) (\gamma^t , \delta)}_{((\Phi^l)^t,0) (0,\Phi^r)}
 \hat {\cal B}^{1/4,2}_{\gamma^t} (q) \hat {\cal B}^{1/4,2}_{\delta^t} (q) \, .
\end{align}
Under the assumption \eqref{assumption}, this expression reduces to
\begin{align}
 s \chi_\Lambda^1 (q) = \chi_0^2 (q) z_F^{(1)} (q)
 \sum_{\Phi^l,\Phi^r}  q^{\frac{|\Lambda^l| + |\Lambda^r| - |\Phi^l| - |\Phi^r|}{4}}
  r_{\Lambda^l \Phi^l} r_{\Lambda^r \Phi^r}
 \hat {\cal B}^{1/4,2}_{(\Phi^l)^t} (q) \hat {\cal B}^{1/4,2}_{(\Phi^r)^t} (q) \, .
\end{align}
Notice that
\begin{align}
 s_\alpha (u^1_{q}) = q^{|\alpha|/4}  \, ,
\end{align}
where we have defined
$u^M_{q}=(q^{1/4}, \ldots , q^{1/4} , 0,0,\ldots | 0, 0 , \ldots)$ with $M$ non-zero
entries. Then we find
\begin{align}
\sum_{\Phi^l} q^{\frac{|\Lambda^l|  - |\Phi^l| }{4}} r_{\Lambda^l \Phi^l}
  \hat {\cal B}^{1/4,2}_{(\Phi^l)^t} (q)
	&= \sum_{\Phi^l, \epsilon , \delta} c^{\Lambda^l}_{\alpha \Phi^l}
 c^{(\Phi^l)^t}_{\epsilon \delta^t} s_\alpha (u^1_{q})
	    \text{sch}_{\epsilon} ({\cal U}_{3/4})
			\text{sch}_{\delta} ({\cal U}_{1/4})  \\
	&= \sum_{\Phi^l, \epsilon , \alpha , \delta} c^{\Lambda^l}_{\delta \Phi^l}
 c^{\Phi^l}_{\epsilon^t \alpha} s_\alpha (u^1_{q})
	    \text{sch}_{\epsilon} ({\cal U}_{3/4})
			\text{sch}_{\delta} ({\cal U}_{1/4}) \nonumber \\
				&= \sum_{\Phi^l , \delta} c^{\Lambda^l}_{\delta \Phi^l}
	    \text{sch}_{\Phi^l} ({\cal U}_{1/4})
			\text{sch}_{\delta} ({\cal U}_{1/4}) = {\cal B}^{1/4 ,2}_{\Lambda^l} (q) \, ,\nonumber
\end{align}
which leads to \eqref{schiM} with $M=1$.

For  generic $M \neq 1$, we can easily arrive at
\begin{align}
 s \chi_\Lambda^M (q) = \chi_0^{2M} (q) ( z_F^{(1)} (q) )^{M^2}
 \sum_{\Phi^l,\Phi^r}  q^{\frac{|\Lambda^l| + |\Lambda^r| - |\Phi^l| - |\Phi^r|}{4}}
  r_{\Lambda^l \Phi^l} r_{\Lambda^r \Phi^r}
 \hat {\cal B}^{1/4,2M}_{(\Phi^l)^t} (q) \hat {\cal B}^{1/4,2M}_{(\Phi^r)^t} (q)
\end{align}
with
\begin{align}
 \hat {\cal B}^{1/4,2M}_{\gamma^t} (q) 	
 = \sum_{\epsilon , \delta} c^{\gamma^t}_{\epsilon \delta^t}
	    {\cal B}^{3/4,M}_{\epsilon} (q)
			{\cal B}^{1/4,M}_{\delta} (q) \, .
\end{align}
Using
\begin{align}
 r_{\alpha \beta} = \sum_{\gamma} c^\alpha_{\beta \gamma} s_\gamma (u^M_1) \, ,\quad
 q^{|\alpha|/4} s_\alpha (u^M_{1}) = s_\alpha (u^M_{q}) =
 \sum_{\alpha_1 , \ldots , \alpha_M} c^\alpha_{\alpha_1  \ldots  \alpha_M}
 \prod_{A=1}^M s_{\alpha_A} (u^1_{q})   \, ,
\end{align}
we obtain
\begin{align}
\sum_{\Phi^l} q^{\frac{|\Lambda^l|  - |\Phi^l| }{4}} r_{\Lambda^l \Phi^l}
  \hat {\cal B}^{1/4,2M}_{(\Phi^l)^t} (q)
	&= \sum_{\Phi^l, \epsilon , \alpha ,\delta} c^{\Lambda^l}_{\delta \Phi^l}
 c^{\Phi^l}_{\epsilon^t \alpha}  s_\alpha (u^M_{q})
	    {\cal B}^{3/4,M}_{\epsilon} (q)
			{\cal B}^{1/4,M}_{\delta} (q)  \\
				&= \sum_{\Phi^l , \delta} c^{\Lambda^l}_{\delta \Phi^l}
	    {\cal B}^{1/4,M}_{\Phi^l} (q)
			{\cal B}^{1/4,M}_{\delta} (q)
			 = {\cal B}^{1/4 ,2M}_{\Lambda^l} (q) \, ,\nonumber
\end{align}
thus we finally have \eqref{schiM} also for $M >1$.

\section{Symmetry generators of coset models}
\label{sec:sgcm}

\subsection{Conventions for su$(M)$ currents}
\label{suM}

With the decomposition \eqref{suMdec}
\begin{align}
 \text{su}(N+M) = \text{su}(N) \oplus \text{su}(M) \oplus \text{u}(1) \oplus ( N , \bar{M} ) \oplus
 ( \bar {N} , M )
\end{align}
the generators can be written as
$(t^\alpha , t^\rho , t^{\text{u}(1)}, t^{(a \bar \imath )} , t^{(\bar a i)})$.
We adopt the following normalization as
\begin{align}
 \text{tr} (t^\alpha t^\beta ) = \delta^{\alpha \beta} \, , \quad
 \text{tr} (t^\rho t^\sigma ) = \delta^{\rho \sigma} \, , \quad
 \text{tr} (t^{\text{u}(1)} t^{\text{u}(1)} )= 1 \, , \quad
 \text{tr} (t^{(a \bar \imath )} t^{(\bar a i )}) =  \delta^{a \bar a} \delta^{\bar \imath i} \, ,
 \label{suMgen}
\end{align}
where the trace is taken over the fundamental representation of $\text{su}(N+M)$.
Using $(e_{IJ})_{KL} = \delta_{IK} \delta_{JL}$, we may express
\begin{align}
&t^{\text{u}(1)} = \frac{1}{\sqrt{NM(N+M)}} \left( M\sum_{i=1}^N e_{ii} - N \sum_{j=1}^M e_{(N+j)(N+j)} \right) \, ,\\
&t^{(a \bar \imath)} =  e_{a (N+\bar \imath)} \, , \quad
t^{(\bar a i)} =  e_{(N+i) \bar a} \, .
\end{align}
The nontrivial structure constants are given by
\begin{align}
& i f^{\alpha \beta \gamma} \, , \quad  i f^{\rho \sigma \tau } \, , \quad
 i f^{\text{u}(1) (a \bar \imath ) ( \bar b j )} = \sqrt{\frac{N+M}{NM}}  \delta^{a \bar b} \delta^{\bar \imath j}\, , \\
& i f^{\alpha ( a \bar \imath ) ( \bar b j )}  = \delta^{\bar b b}t^\alpha_{b \bar a} \delta^{\bar a a} \delta^{\bar \imath j}\, , \quad
i f^{\rho ( a \bar \imath ) ( \bar b j)}  = - \delta^{a \bar b} t^\rho_{i \bar \jmath} \delta^{\bar \imath i}
\delta^{\bar \jmath j}\, .
\end{align}
For the computation of operator products, it is useful to use the following formulas
\begin{align}
& \sum_{B,C} f^{ABC} f^{DBC} = 2 (N+M) \delta^{AD} \, ,  \label{formula1} \\
& \sum_C f^{ABC} f^{CDE} + \sum_{C} f^{DBC} f^{CEA} + \sum_C f^{EBC} f^{CAD} = 0 \label{formula2}
\end{align}
on the structure constants of su$(N+M)$  and
\begin{align}
 \sum_\alpha t^\alpha_{a \bar b} t^\alpha_{c \bar d} = \delta_{a \bar d} \delta_{c \bar b } - \frac{1}{N}
 \delta_{a \bar b} \delta_{c \bar d}  \label{formula3}
\end{align}
on the generators of su$(N)$.

\subsection{${\cal N}=1$ superconformal generator of the dual coset}
\label{App:Ahn}

The coset \eqref{dcoset} with $k=N$, that is
\begin{align}
 \frac{\text{su}(N)_N \oplus \text{su}(N)_M}{\text{su}(N)_{N+M}} \, ,
 \label{dcoset0}
\end{align}
has the central charge
\begin{align}
 c_D = \frac{(N^2 - 1)}{2} \frac{M(3N + M)}{(N+M)(2N+M)} \, ,
 \label{centrald}
\end{align}
and is known to have ${\cal N}=1$ enhanced supersymmetry \cite{Goddard:1986ee,Douglas:1987cv}.
The ${\cal N}=1$ superconformal current is explicitly constructed in \cite{Ahn:2012bp,Ahn:2013ota}.
Here we summarize their results.

In the numerator of \eqref{dcoset0}, the sector of $\text{su}(N)_N$ can be described by free fermions $\Psi^\alpha$ $(\alpha = 1,2, \ldots ,N^2 -1)$, and the su$(N)_N$ currents are given by $J_1^\alpha$ in \eqref{cosetMcurrents}.
The $\text{su}(N)_M$ currents in the numerator are generated by $K^\alpha$ with OPEs
\begin{align}
 K^\alpha (z) K^\beta (w) \sim \frac{M \delta^{\alpha \beta}}{(z - w)^2} + \frac{\sum_{\gamma} i f^{\alpha \beta \gamma} K^\gamma (w)}{z-w} \, .
 \end{align}
Then the sector of $\text{su}(N)_{N+M}$ in the denominator is generated by $J^\alpha_1 + K^\alpha$.
The energy momentum tensor is given in terms of  the Sugawara operators as
\begin{align}
  T_D &= T^{\text{su}(N)}_{N} + T^{\text{su}(N)}_{M} - T^{\text{su}(N)}_{N+M}  \label{AhnEM} \\
    & = \frac{1}{4 N} \sum_{\alpha} J^\alpha_1 J^\alpha_1  + \frac{1}{2(N+M)} \sum_{\alpha} K^\alpha K^\alpha
  - \frac{1}{2(2N+M)} \sum_{\alpha} (J^\alpha_1 + K^\alpha ) (J^\alpha_1 + K^\alpha)    \, .
 \nonumber
 \end{align}
The ${\cal N}=1$ superconformal generator is constructed as \cite{Ahn:2012bp,Ahn:2013ota}
\begin{align}
 G_D = \tilde C \sum_\alpha \left(\Psi^\alpha  J^\alpha _1 - \frac{3N}{M} \Psi^\alpha K^\alpha \right) \, , \quad
 \tilde C = \frac{M}{3\sqrt{N (N+M)(2N+M)}} \, .
 \label{scgd}
\end{align}
The relative coefficient is fixed such that the generator has a regular OPE with the su$(N)_{N+M}$
current in the denominator of \eqref{dcoset0}.
The overall coefficient $\tilde C$ is set to have
\begin{align}
 G_D (z ) G_D (w) \sim \frac{2 c_D / 3 }{(z-w)^2} + \frac{2 T_D (w)}{z-w} \, ,
 \label{N=1OPE}
\end{align}
which is the OPE for ${\cal N}=1$ superconformal generator.

\subsection{${\cal N}=3$ superconformal algebra}
\label{N=3}

The ${\cal N}=3$ superconformal algebra is generated by the spin 2 energy momentum
tensor $T(z)$, three spin $3/2$ super currents $G^{a} (z)$,
three spin 1 currents $J^a (z)$ and a spin $1/2$ fermion $\Psi (z)$.
The OPEs among them are (see, e.g., \cite{Goddard:1988wv})
\begin{align}
 & T(z) T(w) \sim \frac{c/2}{(z - w)^4} + \frac{2 T(w)}{(z-w)^2} + \frac{\partial T(w)}{z-w } \, , \\
 & G^a (z) G^b (w) \sim \delta^{ab} \left( \frac{2c/3 }{(z-w)^3}
+ \frac{2 T(w)}{z-w}
\right)
 + i\epsilon_{abc}
 \left(\frac{2 J^c (w)}{(z-w)^2} + \frac{\partial J^c (w)}{z-w}\right) \, ,  \\
 & J^a (z) J^b (w) \sim  \frac{k \delta^{ab}}{(z -w)^2} + \frac{i \epsilon_{abc} J^c (w)}{z-w} \, , \\
 & J^a (z) G^b (w) \sim  \frac{\sqrt{k} \delta^{ab} \Psi(w)}{(z -w)^2} + \frac{i \epsilon_{abc} G^c (w)}{z-w}\, , \\
 & \Psi (z) G^a (w) \sim \frac{1/\sqrt{k} J^a (w)}{z-w} \, , \\
 & \Psi(z) \Psi(w) \sim \frac{1}{z-w} \, .
\end{align}

\subsection{OPE between an ${\cal N}=3$ superconformal generator}
\label{sec:G3}

We would like to show that the generator $G^3$ in \eqref{G3} satisfies the
operator product \eqref{GGope}.
Through this paper, we implicitly use the normal ordering as
\begin{align}
 (AB) (w) = \frac{1}{2 \pi i } \oint_w \frac{dx}{x-w} A(x )B(w)  \, ,
\end{align}
when two operators are inserted at the same point.
Then we have
\begin{align}
&  \sum_\alpha ( \Psi^\alpha (J_2 ^\alpha - j^\alpha) )(z) \sum_\beta ( \Psi^\beta (J_2 ^\beta - j^\beta) )(w)
 \sim \frac{2 M (N^2 - 1)}{(z-w)^3} \\
& + \frac{1}{z-w} \sum_\alpha \left[  - 2 (J_1^\alpha J_2^\alpha) - 2 (J_1^\alpha j^\alpha) +(J_2^\alpha J^\alpha_2)
  + (j^\alpha j^\alpha) + 2 M ( \partial \Psi^\alpha \Psi^\alpha ) - 2 (J_2^\alpha j^\alpha)\right] (w) \nonumber
\end{align}
and
\begin{align}
& \sum_\rho ( \Psi^\rho (J_2 ^\rho - j^\rho) )(z) \sum_\sigma ( \Psi^\sigma (J_2 ^\sigma - j^\sigma) )(w)
 \sim \frac{2 N (M^2 - 1)}{(z-w)^3} \\
& + \frac{1}{z-w} \sum_\rho \left[  - 2 (J_1^\rho J_2^\rho) - 2 (J_1^\rho j^\rho) +(J_2^\rho J^\rho_2)
  + (j^\rho j^\rho) + 2 N ( \partial \Psi^\rho \Psi^\rho ) - 2 (J_2^\rho j^\rho)\right] (w) \, . \nonumber
\end{align}
Moreover we find
\begin{align}
& ( \Psi^{\text{u}(1)} (\hat J^{\text{u}(1)}- j^{\text{u}(1)}) )(z)   ( \Psi^{\text{u}(1)} (\hat J^{\text{u}(1)}- j^{\text{u}(1)}) )(w)
 \sim \frac{2 N M}{(z-w)^3} \\
& + \frac{1}{z-w} \left[  (\hat J^{\text{u}(1)}\hat J^{\text{u}(1)})
  + ( j^{\text{u}(1)} j^{\text{u}(1)}) + 2 N M ( \partial \Psi^{\text{u}(1)} \Psi^{\text{u}(1)} ) - 2 (\hat J^{\text{u}(1)} j^{\text{u}(1)} )\right] (w) \, . \nonumber
\end{align}

It is useful to rewrite the energy momentum tensor in \eqref{N=3Sugawara} as
\begin{align}
 T = & \frac{1}{4(N+M)} \left[ 2 \delta_{a \bar b} \delta_{\bar \imath j} ( J^{(a \bar \imath )}_2 J^{(\bar b j)}_2 )
  - 2 \sum_\alpha ( J^\alpha j^\alpha )
- 2 \sum_\rho ( J^\rho j^\rho )
- \frac{2}{NM} ( \tilde J^{\text{u}(1)} j^{\text{u}(1)} ) \right]  \nonumber  \\  &- \frac12 ( \Psi^{\text{u}(1)} \partial
\Psi^{\text{u}(1)} )
 - \frac{ \delta_{a \bar b} \delta_{\bar \imath j} }{4NM} \left [ ( \Psi^{(a \bar \imath )} \partial \Psi^{ ( \bar b j ) } ) -  ( \partial \Psi^{ ( a \bar \imath )} \Psi^{ (\bar b j )} ) \right ] \label{T1}
 \\ &
  - \frac {\delta_{a \bar b} \delta_{\bar \imath j} }{4} \left[( \psi^{ ( a \bar \imath ) } \partial \psi^{ (\bar b j ) } ) - ( \partial \psi^{ ( a \bar \imath ) } \psi^{ ( \bar b j ) } ) \right] \, ,  \nonumber
  \end{align}
where we have defined as
\begin{align}
& J^{ ( a \bar \imath )}_2 = \sum_\alpha
 \Psi^\alpha \Psi^{(b \bar \imath )} t^\alpha_{b \bar a} \delta^{\bar a a} - \sum_\rho \Psi^\rho \delta^{\bar \imath i} t^\rho_{i \bar \jmath}\Psi^{ ( a \bar \jmath )}\, ,  \\
& J^{(\bar a i ) }_2 = - \sum_\alpha
\delta^{\bar a a} \Psi^\alpha t^\alpha_{a \bar b} \Psi^{(\bar b i )} +  \sum_\rho \Psi^\rho \Psi^{(\bar a j )} t^\rho_{j \bar \imath} \delta^{\bar \imath i} \, . \nonumber
\end{align}
The formula in \eqref{peter} is useful to arrive at this expression.
With the following identity as
\begin{align}
\delta_{a \bar b} \delta_{\bar \imath j}(J_2 ^{ ( a \bar \imath ) } J_2^{ ( \bar b j )})
 &= - \sum_\alpha \left[ (J^\alpha_1 J^\alpha_2)  +  M ( \Psi^\alpha \partial \Psi^{\alpha} )  \right]
 -  \sum_\rho \left[ (J^\rho_1 J^\rho_2) + N ( \Psi^\rho \partial \Psi^{\rho} ) \right] \nonumber \\
& - \frac{ ( N - \frac1N ) + ( M - \frac1M )  }{2} \delta_{a \bar b} \delta_{\bar \imath j} [  ( \Psi^{ ( a \bar \imath )  } \partial \Psi^{ ( \bar b j ) } )
 - ( \partial \Psi^{ ( a \bar \imath ) } \Psi^{ ( \bar b j ) } ) ]
 \, ,
\end{align}
the expression \eqref{T1} can be reduced to
\begin{align}
 T & = \frac{1}{4(N+M)} \sum_\alpha  \left[ - 2  ( J_1^\alpha J_2^\alpha)
- 2 M  ( \Psi^\alpha \partial \Psi^\alpha )
 - 2 (J^\alpha j^\alpha)
   + (J_2^\alpha J_2^\alpha) + (j^\alpha j^\alpha)  \right] \nonumber \\
&  + \frac{1}{4(N+M)} \sum_\rho \left[  - 2 ( J_1^\rho J_2^\rho)
 - 2 N ( \Psi^\rho \partial \Psi^\rho ) - 2 (J^\rho  j^\rho)
 + (J_2^\rho J_2^\rho)  + (j^\rho j^\rho) \right] \\
  & - \frac{1}{4NM} \left[ 2( \hat J^{\text{u}(1)}   j^{\text{u}(1)} ) - ( \hat J^{\text{u}(1)}   \hat J^{\text{u}(1)} ) - ( j^{\text{u}(1)}   j^{\text{u}(1)} ) \right]  - \frac12  ( \Psi^{\text{u}(1)} \partial
\Psi^{\text{u}(1)} ) \, .
\nonumber
\end{align}
With the above equation we can show that the operator product of $G^3$ in \eqref{G3} is given by \eqref{GGope}.

%\bibliographystyle{JHEP}
%\bibliography{AdS3}

\providecommand{\href}[2]{#2}\begingroup\raggedright\endgroup

\end{document}